\documentclass{aa}  

\usepackage{graphicx}
\usepackage{txfonts}
\usepackage{amsmath}	
\usepackage{amssymb}
\usepackage{xcolor}

\graphicspath{{figs/}}
\newcommand{\comment}[1]{} 
\usepackage[colorlinks=false]{hyperref}
\begin{document}

   \title{Black hole outflows initiated by a large-scale magnetic field}

   \subtitle{Effects of aligned and 
   misaligned 
   spin}


   \author{Bestin James
          \inst{1}\thanks{E-mail: bestin@cft.edu.pl}
        \and
          Agnieszka Janiuk\inst{1}
                   \and
          Vladim\'ir Karas\inst{2}
          }

   \institute{Center for Theoretical Physics, Polish Academy of Sciences, Al. Lotnik\'ow 32/46, 02-668 Warsaw,                  Poland
         \and
             Astronomical Institute, Czech Academy of Sciences, Bo\v{c}n\'{i} II, CZ-141 00 Prague, Czech Republic
             }

   \date{Received --; accepted --}

 
  \abstract
   {Accreting black hole sources show variable outflows at different mass scales. For instance, in the case of galactic nuclei, our own galactic center Sgr A* exhibits flares and outbursts in the X-ray and infrared bands. Recent studies 
   suggest that the inner magnetospheres of these sources have a pronounced effect on such emissions.  
   }
   {Accreting plasma carries the frozen-in magnetic flux along with it down to the black hole horizon. During the in-fall, the magnetic field intensifies and it can lead to a magnetically arrested state. We investigate the competing effects of inflows at the black hole horizon and the outflows developed in the accreting plasma due to the action of magnetic field in the inner magnetosphere and their implications.}
   {We start with a spherically symmetric Bondi-type inflow and introduce the magnetic field. In order to understand the influence of the initial configuration, we start the computations with an aligned magnetic field with respect to the black hole rotation axis. Then we proceed to the case of magnetic fields inclined to the black hole rotation axis. We employ the 2D and 3D versions of \texttt{HARM} code for the aligned field models while using the 3D version for the inclined field and compare the results of computations against each other.}
   {We observe how the magnetic lines of force start accreting with the plasma while an equatorial intermittent outflow develops and goes on pushing some material away from the black hole partially along the equatorial plane, and partly ejecting it out of the plane in the vertical direction. In consequence, the accretion rate also fluctuates.
    The black hole spin direction prevails at later stages and it determines the flow geometry near the event horizon, whereas on larger scales the flow geometry stays influenced by the initial inclination of the field.}
   {}


   \maketitle
%
%
%
\section{Introduction}\label{intro}
Various observational studies point towards the idea that an accreting black hole acts as an engine that powers high-energy phenomena in jets from active galactic nuclei (AGN), including gamma-ray bursts \citep[e.g.,][]{2009herb.book.....D,2012bhae.book.....M}.
Black holes attract the gaseous material from its surrounding environment \citep{ReesVolonteri2007IAUS}. The accretion of plasma onto the central compact object due to its gravitational force produces various observable effects including the relativistic jets and outflows in the form of winds \citep[in particular in stellar mass black holes; see][]{Narayan2013blackholes,FenderGallo2014review}. It is appropriate to assume that plasma at large-scales is magnetized due to its environment. As the accretion proceeds, a large magnetic pressure can be developed near the black hole horizon and it can act against accretion itself. This sometimes leads to the formation of the Magnetically Arrested Disk \citep[MAD;][]{Narayan_mad2003, Tchekhovskoy_mad2011, EHT_2021ApJ}. The accumulated magnetic flux can inhibit further accretion and it plays a role in the formation of relativistic jets observed from the black hole sources. 

A mechanism which supports the formation and collimation of jets from a Kerr black hole surrounded by magnetic field was originally described by \citet{BZ1977}. According to this mechanism, the jet can extract energy from the rotation of the black hole, which was later demonstrated in GRMHD simulations \citep{Tchekhovskoy_mad2011}. On the other hand, in the case of rapidly spinning black holes, Meissner-like expulsion of magnetic field lines is expected which can adversely affect the efficiency of Blandford-Znajek mechanism \citep{BicakJanis1985, KingLasota1975}. It was later shown by \citet{KomissarovMcKinney2007} and others that such an effect is diminished by the inward force of accretion and the Blandford-Znajek mechanism is often sustained. The black hole does not hold the magnetic field by itself. Once the accretion rate begins to diminish, it is possible that the magnetic field lines can push the clumps of plasma away from the black hole. The magnetically inhibited inflow can eventually transform into an outflow above and below the disk, where the current sheet forms \citep{2020ApJ...900..100R,2022ApJ...924L..32R}.

While the occurrence of the magnetically arrested state is frequently observed in GRMHD simulations of accreting black holes, the development of the system is rather sensitive to the conditions at the outer boundary, dimensionality of the computation, and other assumptions about the flow. \citet{2022A&A...663A.169E,ElMellahetal_2023A&A} reveal a separatrix and current sheet in a centrifugally determined flow where particles can be accelerated along a characteristic inclination angle, which is also seen in the synchrotron emission when the computation includes the associated radiation losses. 
In the latter paper these authors present global 3D general relativistic particle-in-cell (PIC) simulations of a black hole magnetosphere, aiming to replicate the observed hot spots and flares around Sgr A*. Their simulations reveal the formation of magnetic flux ropes and the subsequent dissipation of electromagnetic energy through magnetic reconnection. Furthermore, they show that the synchrotron emission from these flux ropes can reproduce the observed flares, with the hot spot's motion and duration being consistent with the model's predictions. Their study also offers constraints on the spin of Sgr A* by studying the dynamics magnetosphere. \citet{Chashkina2021MNRAS.508.1241} presented a series of numerical simulations in which the outflows are activated from the black hole due to the advection of small scale magnetic fields. They used small scale magnetic field loops and such configurations are shown to potentially support the formation of quasi-striped Blandford-Znajek jets. Their configuration also seem to lead to enhanced dissipation and generation of plasmoids in current sheets formed in the vicinity of the black horizon, which can potentially act as mechanism that powers hard X-ray emissions in many accreting black hole systems.

\citet{Crinquand2022PIC} identify a persistent equatorial current sheet and the flaring activity in a set-up relevant for a very low accretion state \citep[see also][]{2023ApJ...943L..29H}. Further, \citet{Kwan2023ApJ...946L..42K} explore Bondi-type accretion and they demonstrate that even in that system the flow can be magnetically arrested and it can produce episodic jets that are launched predominantly along the common rotation axis of the black hole and the accretion disk. They demonstrated that the magnitude of the initial specific angular momentum of the gas plays a vital role in the formation of a MAD state and in the subsequent launching of powerful jets. They also notice that to have a very powerful and stable relativistic jet, the accretion flow has to sustain a MAD state which requires a nonzero value of specific angular momentum for the gas, even though this threshold value is very small (it corresponds to a circularization radius between $10 r_g$ and $50 r_g$).

\citet{Ressler_2021} had also explored zero angular momentum accretion onto a rapidly rotating black hole and they found that the jet power peaks for an initially tilted magnetic field with respect to the black hole spin axis as this geometry maximizes the magnetic flux across the event horizon. They assumed a uniform magnetic field with different tilt angles, with a uniform initial plasma $\beta$ of 100 and a large value of black hole spin ($a = 0.9375$). In this set-up they noticed the development of polar jets almost regardless of the magnetic tilt angles, driven by a nearly magnetically arrested configuration in the equatorial region of the flow. Recently, \citet{2023MNRAS.521.4277R} performed realistic 3D GRMHD model computations in the context of wind-fed accretion from a set of Wolf-Rayet stars near Sgr~A* supermassive black hole. They observe stochastic behavior of the system with a jet changing its direction with respect to the black hole rotation axis and the MAD state being temporarily suppressed during quiescent periods.

\cite{Curd2023MNRAS.518.3441C} investigated MAD state with their GRRMHD simulations and those models resulted in relativistic jets. Their models showed significant variability in outgoing radiation which they attribute to episodes of magnetic flux eruptions. 
\citet{2023arXiv231204149S} explored the possibility of jetted outflows being assisted by repetitive triggering by orbiting perturbers, such as an intermediate secondary black hole.

In this work, we focus on the structure and rate of plasma outflows mediated by the presence of large-scale magnetic fields \citep[e.g.][]{Wald1974} and the resultant evolution of the plasma and the frozen in magnetic field in the vicinity of the black hole horizon. We examine the competing effects of mass inflows and outflows driven by a large-scale asymptotically uniform magnetic field in the Kerr geometry starting from a spherically symmetric inflow. In the case of magnetic field 
aligned with the black hole spin, the process of accretion carries magnetic field lines along with the plasma while intermittent outflows develop mainly in the equatorial region. 

Here, we use an initially spherically symmetric inflow, similar to the \citet{Bondi1952} solution, as the initial stationary solution upon which the magnetic field is imposed, and we evolve it with time. We also consider 
misalignment of the magnetic axis with respect to the black hole angular momentum vector. In the latter case the outflows are significantly distorted by the rotating black hole and we notice the formation of a low density funnel region nearby the rotation axis. 
Thus we illustrate an important role played by the field inclination. Indeed, in the limit of test particles, it can be seen \citep{KopacekKaras2020ApJ} that the particle acceleration is  significantly enhanced by frame-dragging effects when the magnetic field is moderately inclined, in contrast to the aligned case.

The article is organized as follows. In Section \ref{numerical_setup}, we explain our numerical setup and the models. We describe the specific results from our models in Section \ref{results}, and we discuss them in Sect. \ref{diss}. Finally, we give the conclusions in Section \ref{conclusions}.

\section{Numerical Setup and Models}\label{numerical_setup}
\subsection{Basic equations and the numerical code}
We model the accretion onto a black hole in a fixed Kerr metric.
The metric is not evolved in time with the assumption that the accreted mass is negligible in comparison to the black hole mass over the timescale considered. For other possibilities, e.g. studies with time evolution of Kerr metric and black hole spin and the mass changes due to massive accretion in collapsars, see the study by \cite{Krol2021ApJ...912..132K}.
We use a modified version of the \texttt{HARM} code which is a conservative and shock capturing scheme for evolving the equations of ideal magnetohydrodynamics in general relativity (GRMHD) \citep{Gammie2003,Noble2006,Janiuk_etal_harm_2018}. So the magnetic field lines are frozen into the plasma. The code follows the evolution of the accreting gas and magnetic field by solving the continuity, energy-momentum conservation and induction equations in the GRMHD scheme:
\begin{equation}
    \nabla_\mu(n u^{\mu})= 0,
\end{equation}    
\begin{equation}
    \nabla_\mu(T^{\mu\nu}) = 0,
\end{equation}
\begin{equation}
    \nabla_\mu(u^\nu b^\mu - u^\mu b^\nu) = 0,
\end{equation}
where $\nabla_\mu$ represents the covariant derivative, $n$ is the baryon number density in the fluid frame, $u^{\mu}$ is the four-velocity of the gas, $u$ is the internal energy, $p$ is the gas pressure, and $b^{\mu}$ is the magnetic four-vector. $T$ is the stress-energy tensor which comprises matter and electromagnetic parts, $T^{\mu \nu} = T^{\mu \nu}_{\mathrm{gas}} + T^{\mu \nu}_{\mathrm{EM}}$. These can be computed as

\begin{equation}\label{eqn:T_MA}
    T^{\mu \nu}_{\mathrm{gas}} = (\rho + u + p )u^{\mu}u^{\nu} + pg^{\mu \nu},
\end{equation}

\begin{equation}\label{eqn:T_EMb}
    T^{\mu \nu}_{\mathrm{EM}} = b^2 u^{\mu}u^{\nu} + \frac{1}{2}b^2 g^{\mu \nu} - b^{\mu} b^{\nu},
\end{equation}
where $\rho = m n$ with $m$ the mean rest mass per particle.
The magnetic field $B$ in the observer frame is connected to the magnetic four-vector $b^\mu$ in such a way that $b^i = (B^i + b^tu^i)/u^t$ where $b^t = B^iu^{\mu}g_{i\mu}$.
In the code, we use the Gaussian units with the natural unit convention $G=c=M=1$ and a factor of $1/\sqrt{4\pi}$ is absorbed into the definition of the Faraday tensor $F$. Thus the velocities are dimensionless and the length and time scales can be measured in units of the black hole mass $M$. Therefore, the length will be in units of $r_g = GM/c^2$ and time will be in units of $t_g=GM/c^3$ and thus the black hole mass can scale the simulation results. 

\texttt{HARM} uses a modified version of the spherical Kerr-Schild (KS) coordinates for its internal grid. In the KS coordinates, the accretion of matter proceeds smoothly through the black hole horizon and thus the flow evolution can be tracked properly without encountering coordinate singularities. The grid has been adjusted to be logarithmic in radius near the black hole and super-exponentially spaced in the outer region to achieve finer resolution closer to the black hole. In the polar code coordinates, we also put an option to adjust the grid spacing close to the equatorial plane which can be useful in the evolution of accretion disk flows. In the code, the KS radius $r$ has been replaced by the logarithmic radial coordinate $x^{[1]}$, such that 
\begin{equation}\label{eq:mks1}
    r = e^{x^{[1]}};
\end{equation}
the KS latitude $\theta$ has been substituted by $x^{[2]}$,
\begin{equation}\label{eq:mks2}
    \theta = \pi x^{[2]} + \frac{(1-h)}{2}\sin\left(2\pi x^{[2]}\right),
\end{equation}
and the azimuthal angle $\phi$ is kept the same, i.e.
\begin{equation}\label{eq:mks3}
   \phi = x^{[3]}.  
\end{equation}

The outer boundary of the computational domain is set either at $10^5~r_g$ or $10^4~r_g$ in our models so that there are no considerable effects of it on the accretion flows near the horizon where our study is focused. The inner boundary is located at $0.98~r_{H}$, i.e. a fraction inside the event horizon of the black hole, for the corresponding value of black hole spin $a$ so that the flow is causally connected. The 2D grid domain has a resolution of $600 \times 512$ in the ($r,\theta$) directions. For the 3D models, we use a grid resolution of $288 \times 256 \times 128$ in the $(r, \theta, \phi)$ directions. We use a uniform angular resolution for the grid by setting the $h$ parameter to the value 1.0 in Equation \ref{eq:mks2} since the flow is initially spherically symmetric. 

The adiabatic index is assumed to be that of an ideal gas with $\gamma = 4/3$. We use the plasma $\beta$ parameter which is defined as the gas to magnetic pressure ratio ($\beta\equiv p_\mathrm{gas} /p_\mathrm{mag}$) to initially set either the maximum or minimum magnetic field strength in the domain when the field is enabled, where, $p_\mathrm{gas} = (\gamma - 1)u$ and $p_{\rm mag} = b^2/2$. In the models we use, with the initial spherically symmetric Bondi-type inflows, the maximum internal energy of the gas is located near to the black hole horizon. We normalize the field strength either by the maximum or minimum value of beta in the domain as given in Tables \ref{tab:eq_outflow_models} and \ref{tab:incl_models}. 
The models with $\beta_{max} = 0.1$ in the domain has a minimum $\beta$ value of $ \sim 0.0027$ at $10^4~r_g$ at the initial time (in comparison with the models with inclined field with $\beta_{min} = 0.0003$ at $10^4~r_g$ given in Table \ref{tab:incl_models}). In the models with the aligned field, the distribution of $\beta$ along the equator mainly depends on the gas distribution while in the inclined field models they vary in a different fashion along the equatorial region due to inclined geometry of the field as well. In these models with $\beta_{min} = 0.0003$, the maximum $\beta$ in the domain is $\sim 10^6$ near the horizon and hence it has a smoother distribution as compared to the aligned field models.  

\subsection{The models and the initial set-up}

\begin{figure}[tbh!]
    \centering
    \includegraphics[width=0.48\textwidth]{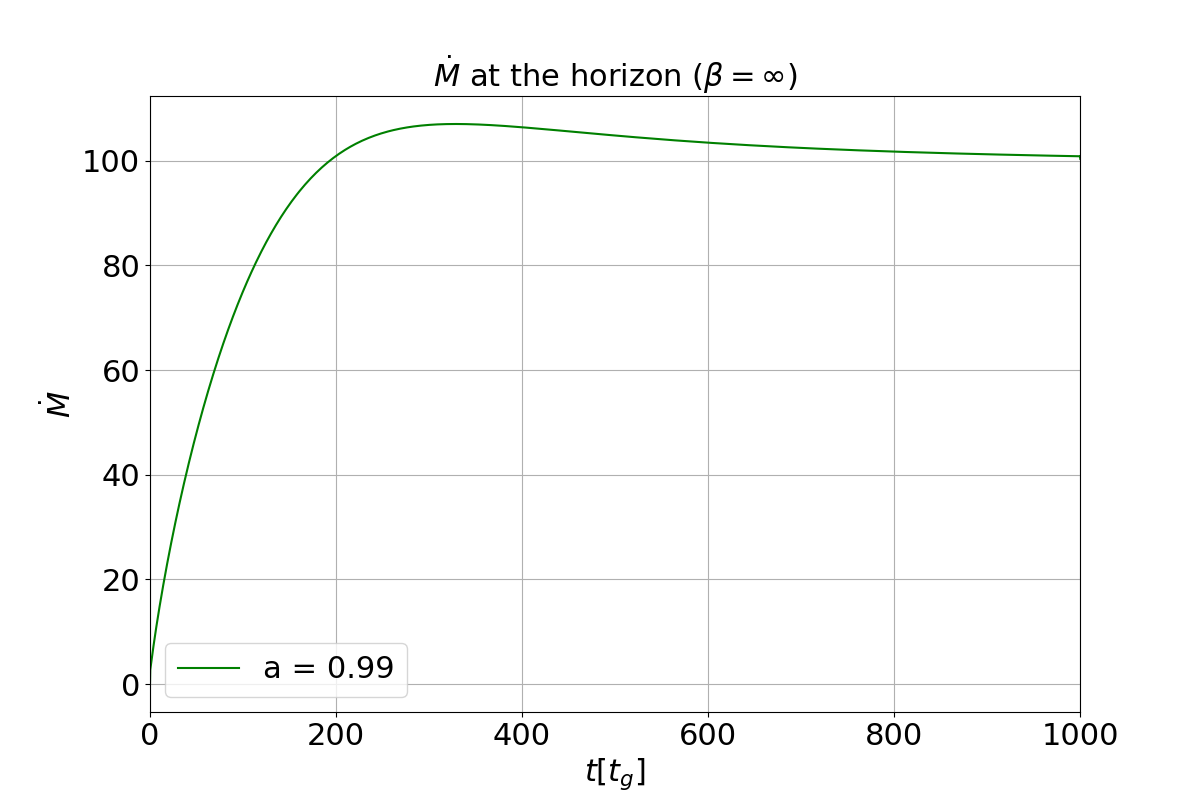}
    \caption{Mass accretion rate at the black hole horizon for the initial part of the simulation before the magnetic field is introduced ($\beta = \infty$). The accretion rate initially grows and saturates to a quasi-steady value as time proceeds. It is given in geometric units which can be scaled to physical units with the black hole mass $M$. The plot is given for only one value of spin, but it is the same for all other values. }
    \label{fig:mdot_beta_infty}
\end{figure}

We use the dimensionless spin parameter $a$ to quantify the black hole rotation, with $a=1$ being the maximally spinning case. We begin our models with a uniform density (of $10^{-1}$ in code units) over the entire computational domain with a Kerr black hole at the center and with zero magnetic field. This configuration is evolved in time, as we see the gas starts accreting to the black hole, and the mass accretion rate rises and reaches a quasi-steady value. The evolution of accretion rate at this stage is plotted in Figure \ref{fig:mdot_beta_infty}. At this point (at $1000~t_g$ given in Fig. \ref{fig:mdot_beta_infty} for our 2D and 3D models), we introduce the magnetic field of the chosen strength in our code. Similarly in our 3D models with the inclined field we introduce, at a similar stage, a magnetic field of relatively lower strength 
not to disrupt the gas flow abruptly due to the inclined field geometry to the BH rotation axis.

\begin{figure*}[tbh!]
    \centering
    \includegraphics[width=0.33\textwidth]{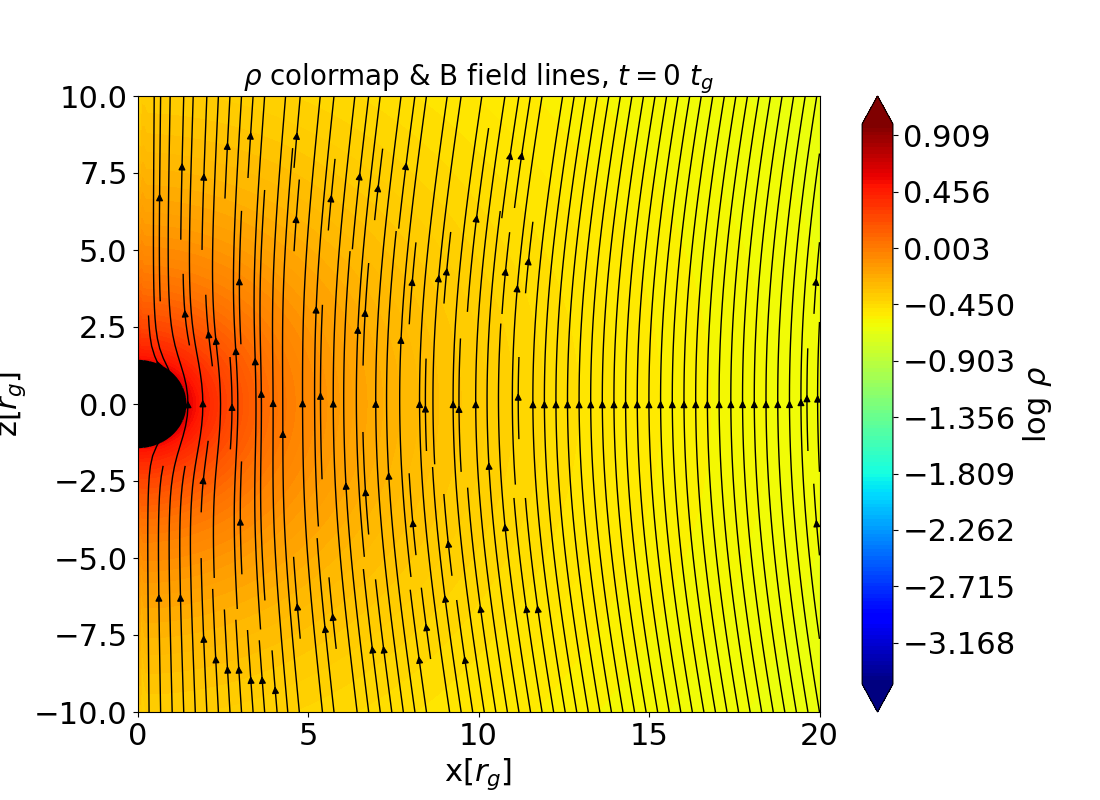}
    \includegraphics[width=0.33\textwidth]{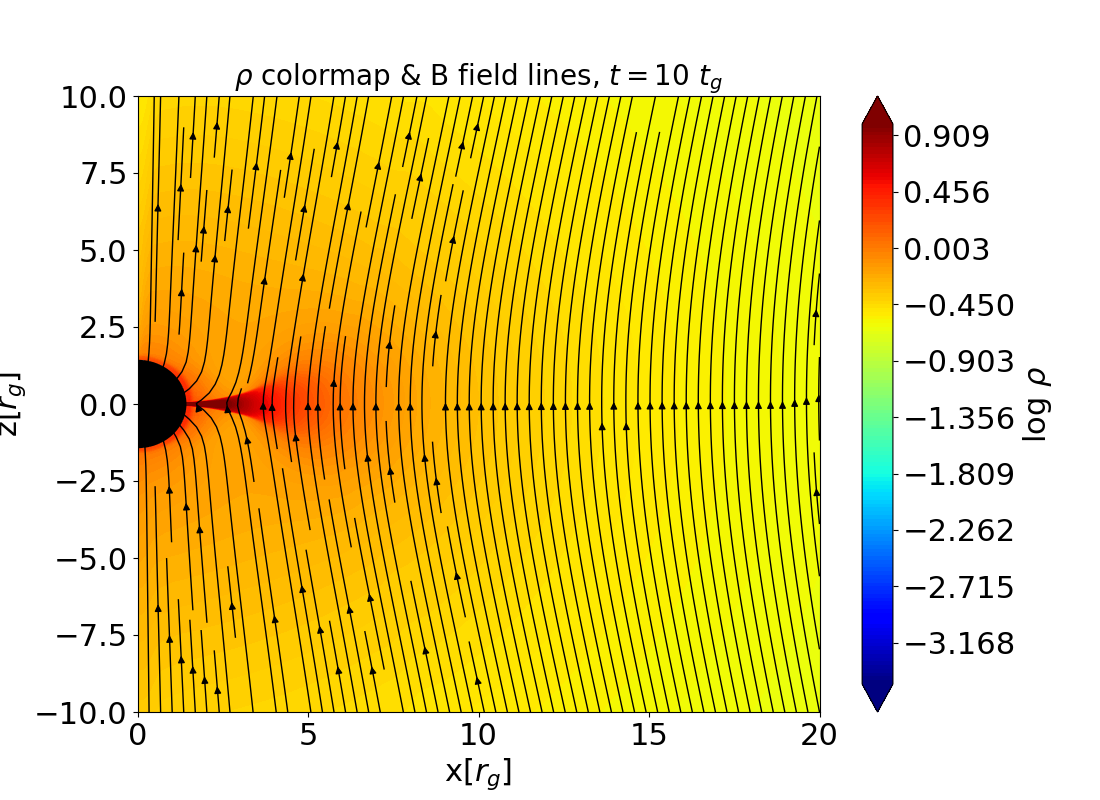}
    \includegraphics[width=0.33\textwidth]{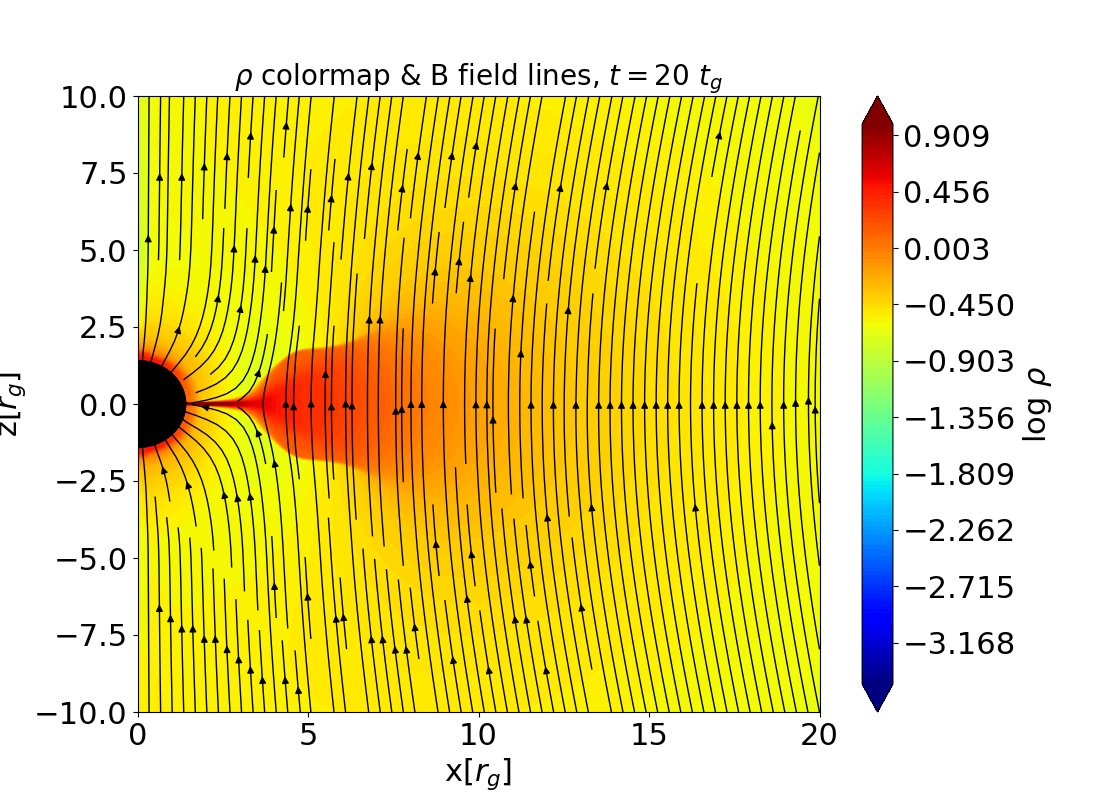}
    \includegraphics[width=0.33\textwidth]{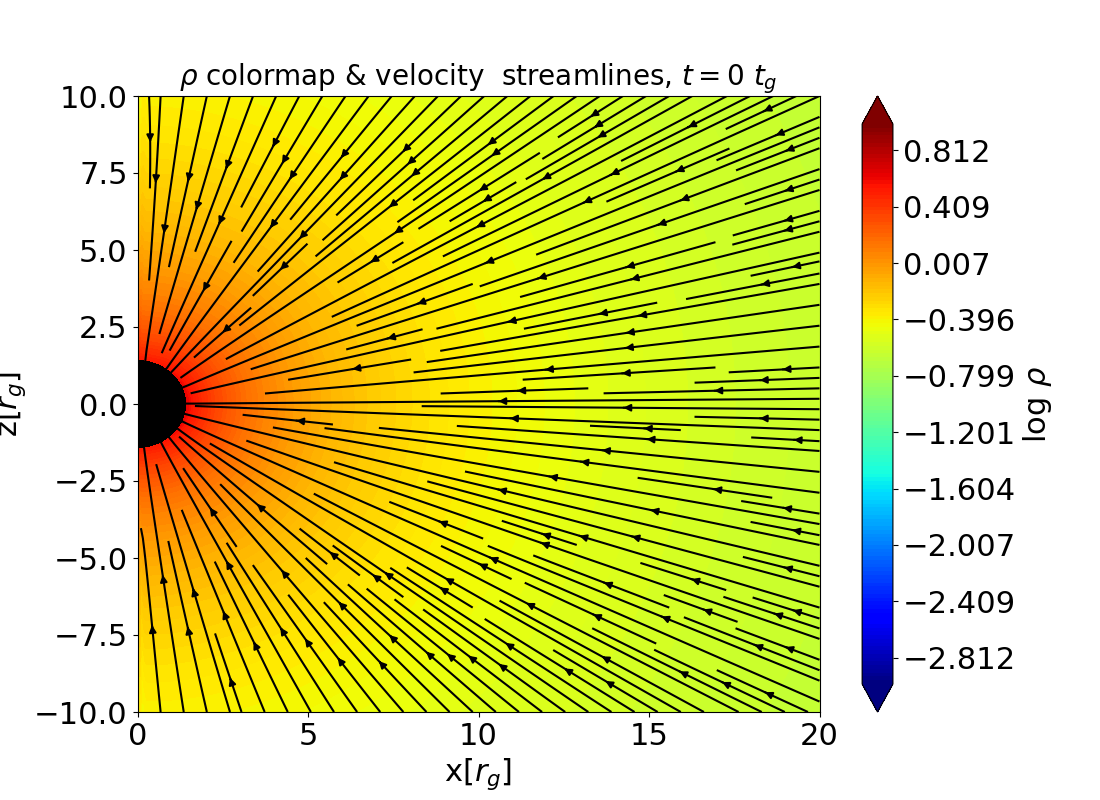}
    \includegraphics[width=0.33\textwidth]{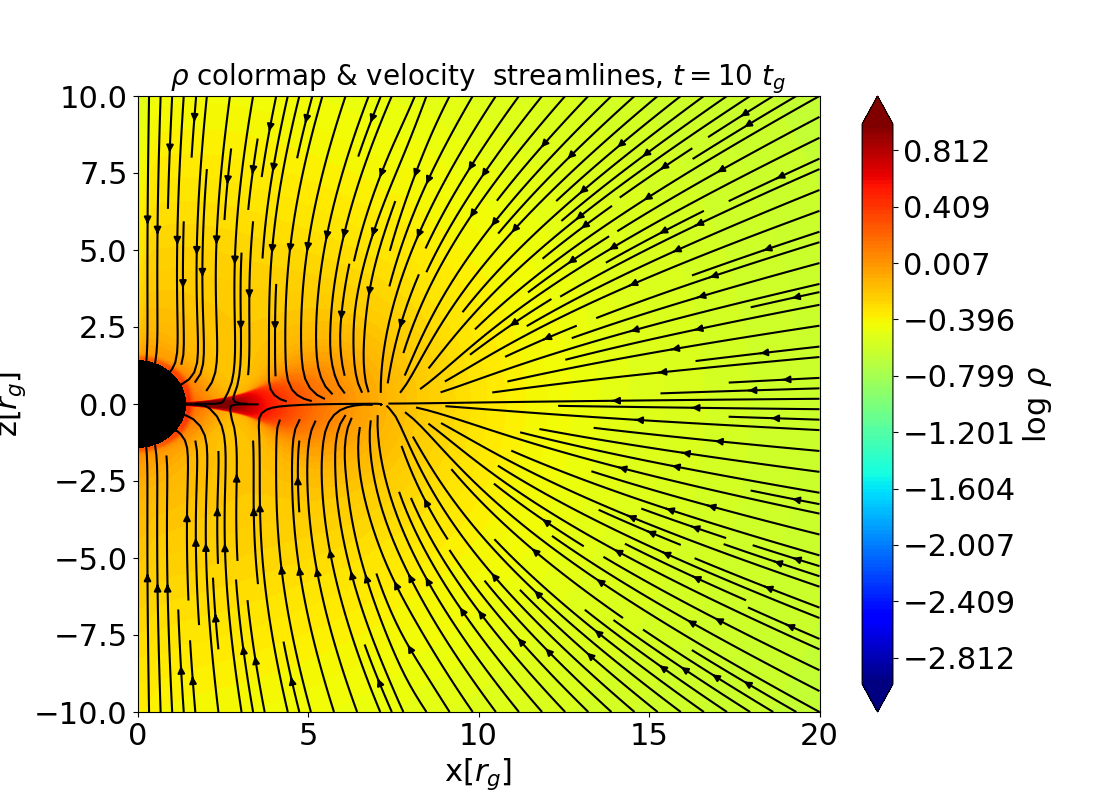}
    \includegraphics[width=0.33\textwidth]{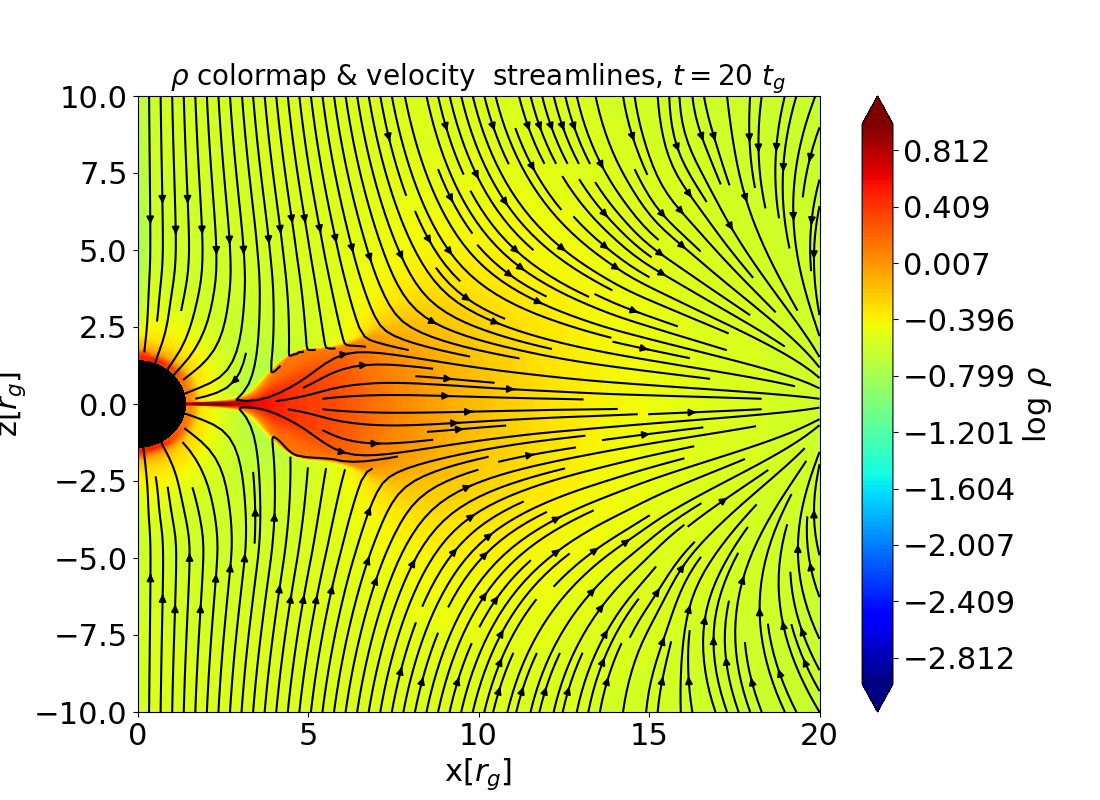}
    \caption{The initial and evolved states of our fiducial model (b01.a90.2D, i.e. with $\beta$ = 0.1 and a = 0.90) after the magnetic field is turned on, with fluid density in color contours with the magnetic field lines plotted on top (top row) and with over-plotted velocity streamlines (bottom row). The top panel demonstrates magnetic reconnection events in the equatorial region while the bottom panel shows the outflows being developed and proceeding outward.}
    \label{fig:[b01_a90_2D]init_density_B}
\end{figure*}

\subsubsection{Magnetic field aligned with the BH rotation axis}

The uniform magnetic field, which is vertically aligned to the rotation axis of the black hole, is prescribed in the code (the \citet{Wald1974} solution). For this configuration, the magnetic field can be fully described by the only non-vanishing components of the vector potential 
\begin{equation}
        A_t = B_0 a \left[r\Sigma^{-1}(1+\cos^2\theta)-1\right],
\end{equation}
\begin{equation}
        A_{\phi} = B_0\left[\textstyle{\frac{1}{2}}(r^2+a^2)-a^2r\Sigma^{-1}(1+\cos^2\theta)\right]\sin^2\theta,
\end{equation}
where $\Sigma(r,\theta) = r^2 + a^2\cos^2\theta$ is a known metric function of coordinates and $B_0$ is a constant with the meaning of magnetic field intensity at large radius.

We start with 2D models with different values of spin, $a=0.99,~0.90$ and $0.60$, so from the near to maximally spin case to a moderate spin, to understand the effect black hole rotation on the magnetic field and the outflow development. We use a range of $\beta$-parameter within the flow ranging from $0.1$ to $1.0$, so that we cover the magnetic pressure dominated cases as well models in equipartition of the gas and magnetic pressure at the initial stage. Let us note that the highly magnetized case ($\beta = 0.1$) is supposed to exhibit the Meissner-like expulsion of the field lines and its potential influence on the gas flow. 
\citet{Karas_2020ragtime} performed a similar study where they investigated the very initial stages of magnetic field and plasma evolution in the presence of a large-scale asymptotically uniform magnetic field. 
We note here that the original \citet{Wald1974} configuration was determined as an electro-vacuum solution, whereas now we examine its modification at later stages by the presence of matter. Indeed, with a better-resolved inner region ($r\lesssim10$ gravitational radii) we confirm that the initial electro-vacuum structure becomes very quickly cancelled by dragging the magnetic field lines into the black hole event horizon.

The left-most panels in Figure \ref{fig:[b01_a90_2D]init_density_B} show the initial configuration of the model with a black hole spin of $a=0.90$ and initial maximum $\beta=0.90$ in terms of the initial density distribution, the initial geometry of the magnetic field and also the velocity streamlines. 


We also investigate two 3D models for the case of aligned magnetic field for the models with maximum magnetization ($\beta_\mathrm{max} = 0.1$), to investigate the non-axisymmetric effects that can occur in the evolution of the above 2D models and to make sure that they are physically viable. Here, we use the same initial setup as described above where we start from a spherically symmetrically inflow to arrive at a quasi-steady accretion rate. At this point, we introduce a magnetic field given by the \cite{Wald1974} solution and follow the evolution of the system.

\subsubsection{Case of inclined magnetic field}

The initial density distribution for these models are obtained in the same way as for the models described in the previous section. Thus all these models start with a spherically symmetric Bondi-type inflow with a quasi-steady accretion rate.
The magnetic field inclined with the rotation axis of the black hole is prescribed according to the form given by \citet{BicakJanis1985}. 
For this configuration, the components of the magnetic field vector in the observer frame are given as 


\begin{multline}
    \mathcal{B}_\phi = A^{-1/2}(\Sigma + 2 Mr)^{-1/2} \Sigma^{-2}  \{ 4B_0 M^2 r^2 a^3 (1+\cos^2 \theta) \sin\theta \cos\theta \\
    - B_1 a [\Sigma^2(\Sigma + Mr) - 4M^2r^2a^2\sin^2\theta]\cos\psi \\
    - B_1[r\Sigma^3 - Ma^2 \cos^2\theta\Sigma^2 \\
    - 2M^2ra^2\sin^2\theta (r^2 -a^2\cos^2\theta)] \sin \psi \}, 
\end{multline}


\begin{multline}
    \mathcal{B}_\theta = - (\Sigma + 2Mr)^{-1/2} \{ B_0 \sin \theta [r + \Sigma^{-2} a^2M (r^2 - a^2\cos^2\theta)(1+\cos^2)] \\
    -B_1\cos \theta (r\cos \psi - a\sin\psi) + B_1 a M \Sigma^{-2} \cos \theta \\
    \times [\![ a[r^2 (1+\sin^2\theta) + a^4 \cos^4 \theta]\cos \psi \\
    +r[r^2 + a^2(\cos^2 \theta - 2 \sin^2\theta)] \sin \psi]\!] \},
\end{multline}

\begin{multline}
    \mathcal{B}_r = A^{-1/2} \Sigma^{-2} \{ B_0 \cos\theta [\![ (r^2 + a^2) [(r^2 - a^2) (r^2 - a^2 \cos^2 \theta) \\
    + 2a^2 r (r-M)(1+\cos^2\theta)] - a^2\Delta\Sigma \sin^2\theta ]\!] \\
    + B_1 r \Sigma^2\sin\theta [(r-M)\cos\psi - a\sin\psi] + B_1 M\sin\theta \\
    \times [(r^2+a^2)(r^2-a^2\cos^2\theta) - \Sigma a^2\sin^2\theta] (r\cos\psi - a\sin\psi) \}.
\end{multline}


The angle of inclination with respect to the black hole rotation axis is set by the constants $B_0$ and $B_1$ whose values represent the relative field strengths in the $Z$ (polar) and $X$ directions respectively. Thus the angle of inclination is given by $\tan^{-1}(B_1/B_0)$. All the models with magnetic field inclined to the black hole rotation axis are run in 3D since they are non-axisymmetric configurations to begin with. 

\section{Results}\label{results}
\subsection{Models with the aligned magnetic field}

A summary of models we studied are listed in Table \ref{tab:eq_outflow_models} and their names reflect the parameter values used for them and the setup (2D or 3D). 
The middle and right columns of Figure \ref{fig:[b01_a90_2D]init_density_B} show the evolved states of a fiducial model b01.a90.2D up to a radius of $20~r_g$. In regions below $5~r_g$ in the these panels, we can notice frozen in magnetic field lines getting reconnected while being accreted onto the black hole. 

In the beginning of the simulations (at $t = 0~t_g$), the magnetic field lines are expelled from the black hole horizon due to the extreme rotation of the black hole. This effect has been observed previously in GRMHD simulations in the case of more extremally spinning black holes \citep{KomissarovMcKinney2007}. But, as the accretion proceeds according to the initially prescribed Bondi solution, the plasma drags in the magnetic field along with it to the black hole horizon and the field lines start bending in the equatorial region of the flow. As the accretion continues, the idealized initial configuration rapidly evolves to a complex structure with more turbulent field lines in the accreting region near to the black hole and more organized field lines in the funnel like regions close to the black hole's rotation axis. Thus the Meissner-like expulsion of magnetic field lines disappears immediately with the beginning of the accretion flow. It can be seen from our models also that such an expulsion of field lines which weakens the Blandford-Znajek mechanism should diminish rapidly in the case of accreting black holes. 

The density structures in the bottom middle and right panels of Figure \ref{fig:[b01_a90_2D]init_density_B} show the mass outflows developed in the equatorial region and their velocity directions. This is more clear in the velocity streamlines overplotted in the Figure. All our magnetized models with aligned field develop equatorial mass outflows similar to the model depicted in Figure \ref{fig:[b01_a90_2D]init_density_B}.



\begin{table*}[tbh!]

	\centering
	\caption[Summary of aligned field models investigated for the equatorial outflows]{Summary of models with aligned field.} 
	\label{tab:eq_outflow_models}
	\begin{tabular}{lcccccc} 
		\hline
		Model$^\dag$ & $\beta_{max}$ & $a$ & Final $\dot M_{\rm in,H}$& $\langle\phi_{\rm BH}\rangle_t$ & $\langle\dot M_{{\rm out,10r_{\it g}}}\rangle_t$ & $\langle\dot M_{{\rm out,10r_{\it g}}}\rangle_t$\\
            &   &  & (code units) &  & (code units) & ($M_{\odot}$ yr$^{-1}$) \\
		\hline

        b1.a0.2D  & 1.0 & 0    & 4.91 & 43.08 & 3.00 $\times 10^{-5}$& $1.060 \times 10^{-7} $\\
        b1.a60.2D & 1.0 & 0.60 & 0.49 & 41.95 & 3.07 $\times 10^{-5}$& $1.084 \times 10^{-7} $\\
        b1.a90.2D & 1.0 & 0.90 & 0.62 & 35.06 & 4.82 $\times 10^{-5}$& $1.703 \times 10^{-7} $\\
        b1.a99.2D & 1.0 & 0.99 & 1.54 & 28.72 & 8.11 $\times 10^{-5}$& $2.864 \times 10^{-7} $\\

        b05.a0.2D  & 0.5 & 0    & 2.70 & 45.73 & 2.36 $\times 10^{-5}$& $8.325 \times 10^{-8} $\\
        b05.a60.2D & 0.5 & 0.60 & 0.71 & 43.44 & 2.45 $\times 10^{-5}$& $8.666 \times 10^{-8} $\\
        b05.a90.2D & 0.5 & 0.90 & 1.08 & 35.57 & 3.99 $\times 10^{-5}$& $1.410 \times 10^{-7} $\\
        b05.a99.2D & 0.5 & 0.99 & 2.92 & 28.88 & 6.87 $\times 10^{-5}$& $2.429 \times 10^{-7} $\\

        b01.a0.2D  & 0.1 & 0    & 2.77 & 46.78 & 2.16 $\times 10^{-5}$ & $7.645 \times 10^{-8} $\\
        b01.a60.2D & 0.1 & 0.60 & 2.98 & 44.67 & 2.81 $\times 10^{-5}$ & $9.941 \times 10^{-8}$ \\
		b01.a90.2D & 0.1 & 0.90 & 4.61 & 35.68 & 4.29 $\times 10^{-5}$ & $1.515 \times 10^{-7}$ \\
		b01.a99.2D & 0.1 & 0.99 & 12.25 & 29.04 & 9.18 $\times 10^{-5}$& $3.243 \times 10^{-7}$ \\
        \hline
        b01.a90.3D & 0.1 & 0.90 & 3.18 & 43.36 & 8.50 $\times 10^{-5}$ &  $3.003 \times 10^{-7}$\\
        b01.a99.3D & 0.1 & 0.99 & 8.05 & 32.41 & 12.65 $\times 10^{-5}$ &  $4.468 \times 10^{-7}$\\
        
		\hline
	\end{tabular}\\[6pt]
\tablefoot{The models are parameterized by the initial maximum plasma $\beta$ and the dimensionless Kerr parameter $a$. The time-averaged magnetic flux on the horizon and the equatorial mass outflow rate at 10 $r_g$ are computed after the magnetic field is turned on. The outflow rate given in the last column is the estimated value in physical units considering our model of mass ejection with the mass of the M87 central engine. The final accretion rate is taken at $2000~t_g$ after the magnetic field has been turned on.\\[3pt] $^\dag$ Notation of the first column indicates the simulation run, the assumed spin value ($0\leq a<1$), and the adopted 2D vs.\ 3D computation.}
\end{table*}

\begin{figure}[tbh!]
    \centering
    \includegraphics[width=0.47\textwidth]{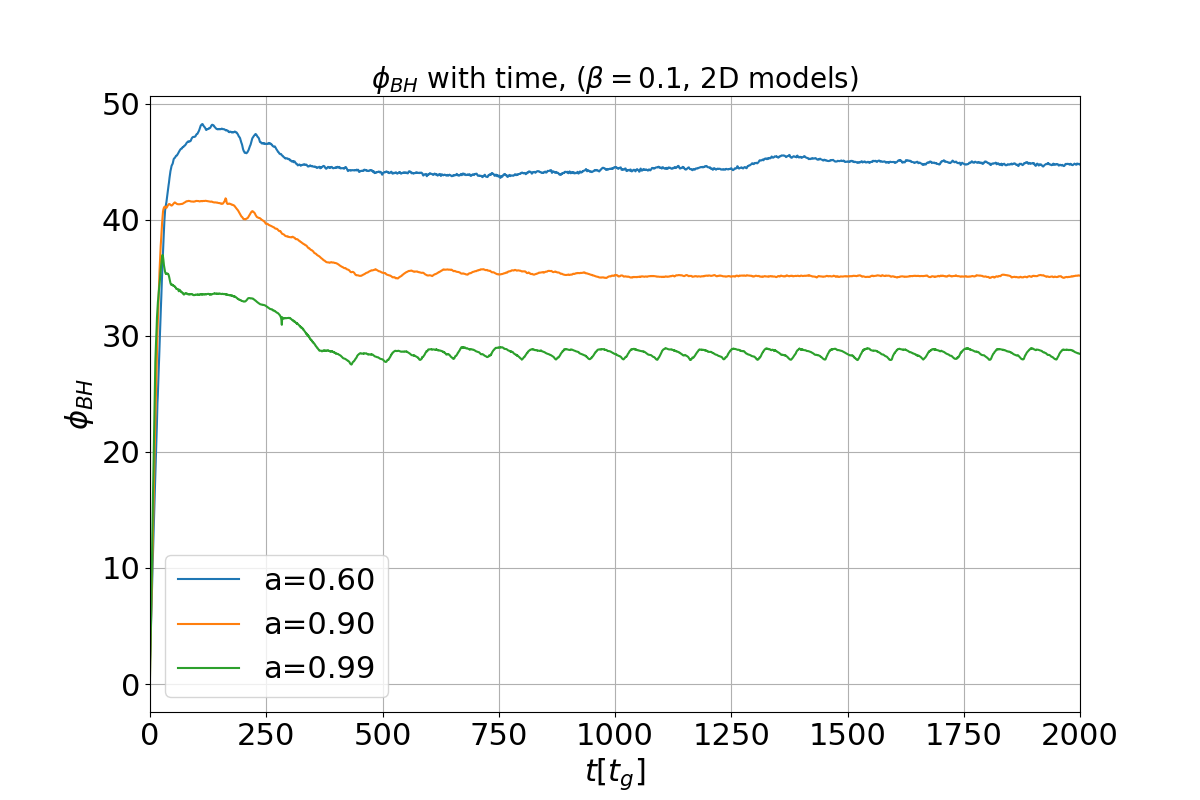}
    \caption{The evolution of magnetic flux on the black hole horizon with time for the 2D models with initial $\beta = 0.1$.}
    \label{fig:[b01_2D]_phiBH_t}
\end{figure}

\begin{figure}[tbh!]
    \centering
    \includegraphics[width=0.47\textwidth]{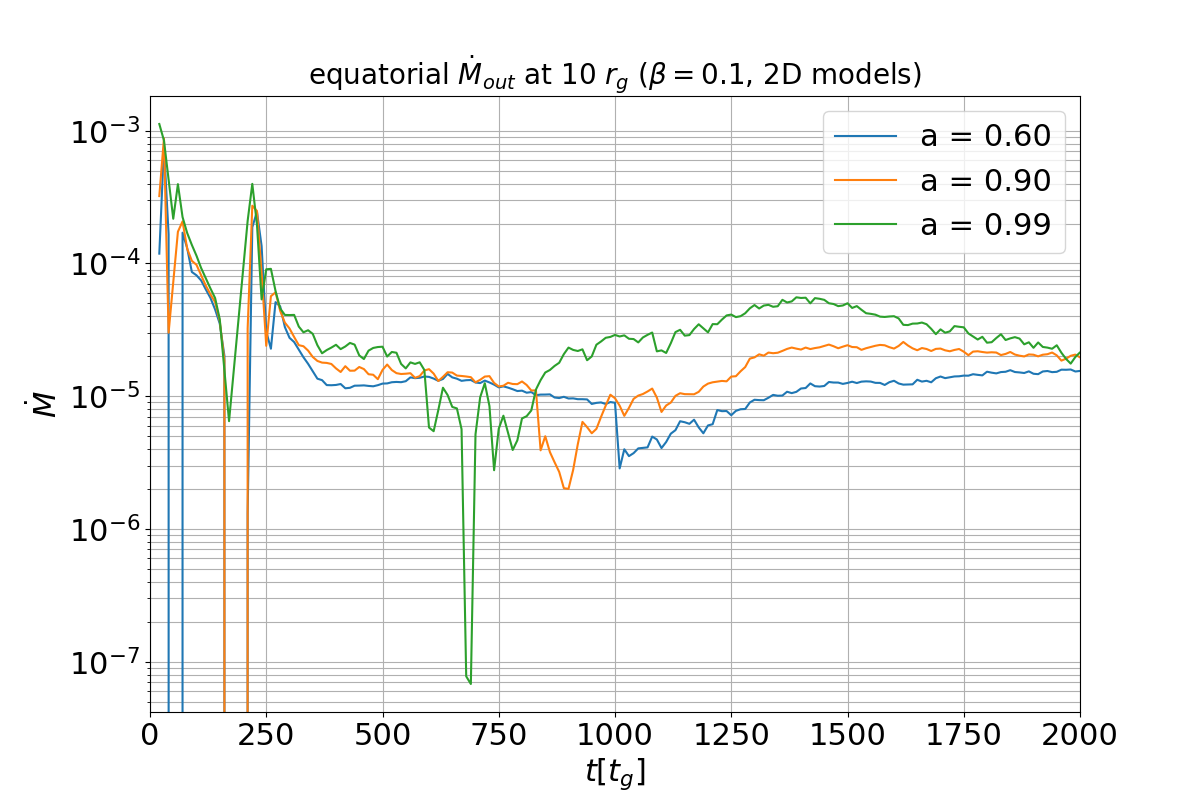}
    \caption[Mass outflow rate with time for the 2D models with $\beta = 0.1$]{The equatorial mass outflow rate with time at 10 $r_g$, after the magnetic field is turned on, for the 2D models with $\beta = 0.1$ and different spin values.}
    \label{fig:[b01_2D]_mout_t}
\end{figure}

\begin{figure}[tbh!]
    \centering
    \includegraphics[width=0.47\textwidth]{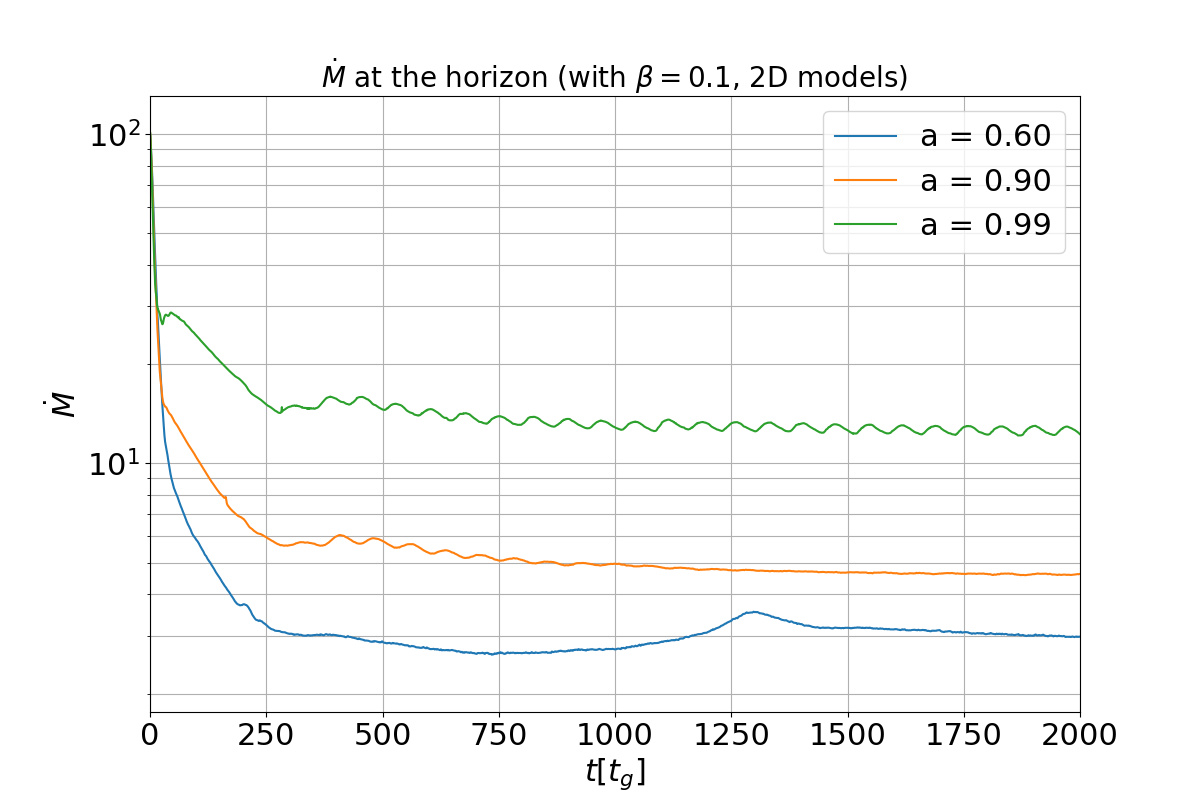}
    \caption{The evolution of inward mass accretion rate at the black hole horizon with time after the magnetic field is turned on for the model with $\beta = 0.1$.}
    \label{fig:[b01_2D]_mdot_t_log}
\end{figure}

\begin{figure*}[tbh!]
    \centering
    \includegraphics[width=0.33\textwidth]{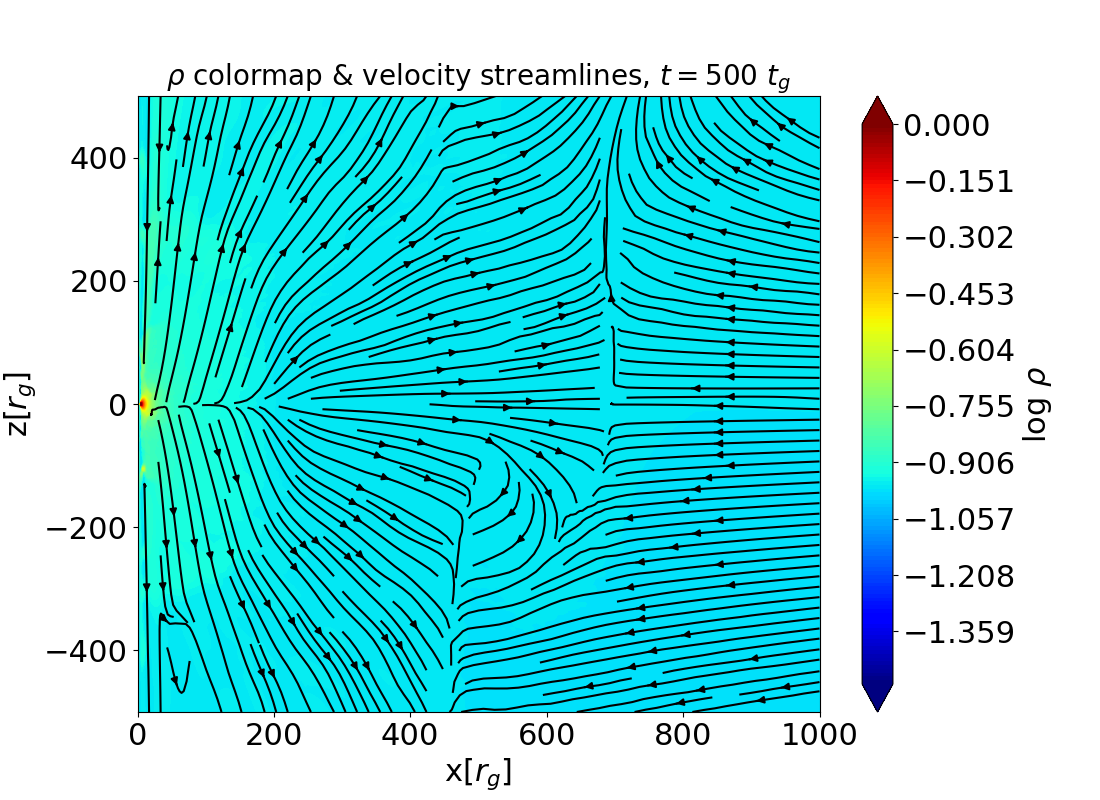}
    \includegraphics[width=0.33\textwidth]{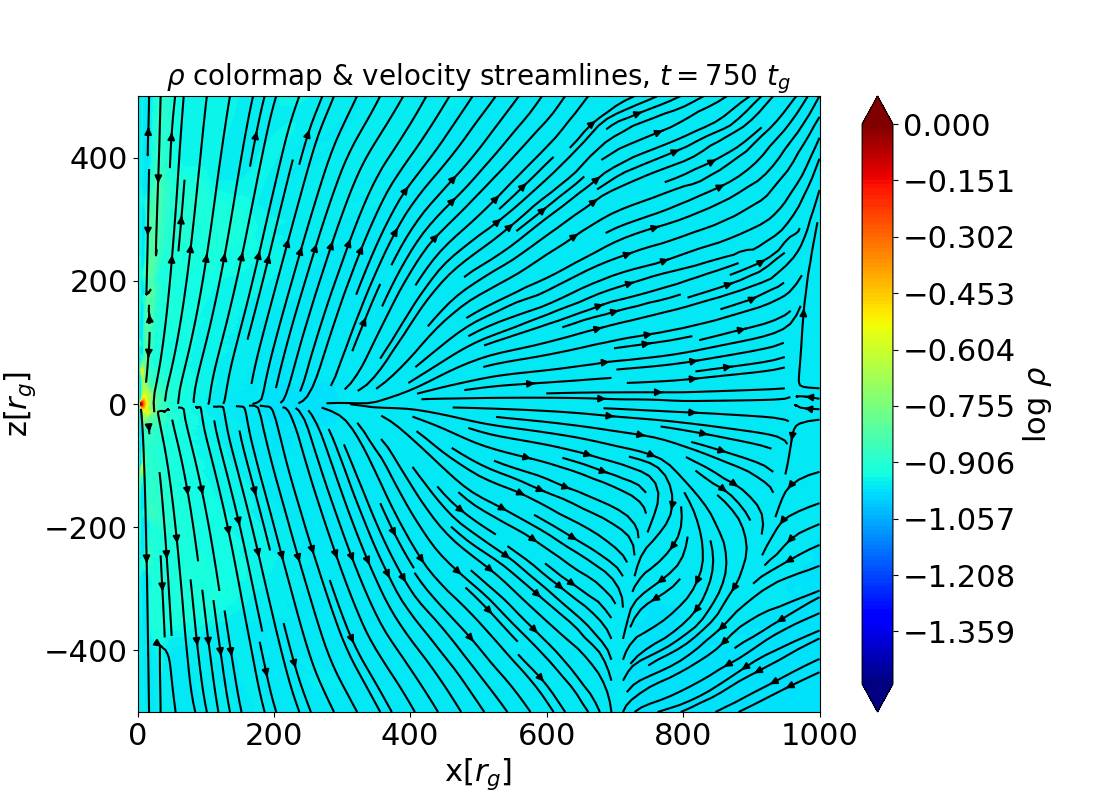}
    \includegraphics[width=0.33\textwidth]{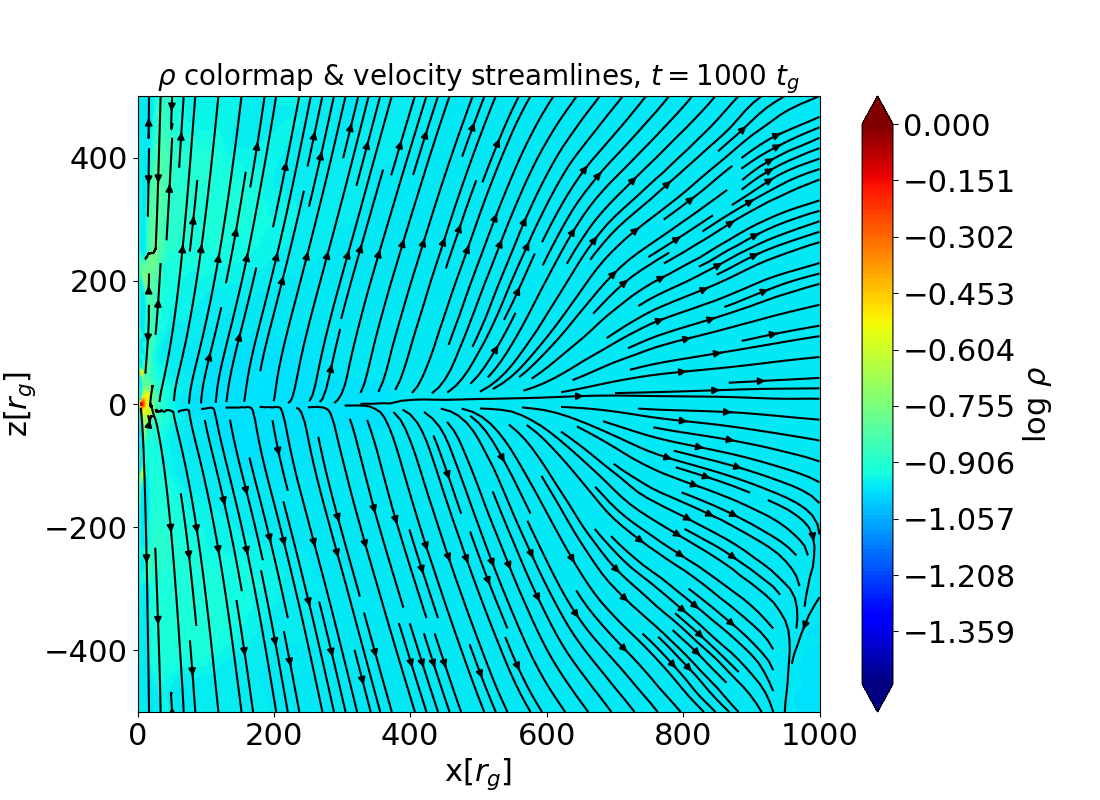}
    \caption{The density color maps over-plotted with the velocity streamlines for one of our representative model b01.a99.2D up to a radius of $1000~r_g$ showing the outflows extending to larger scales (eventually to the outer boundary of the computational domain), at time t = 500, 750 and 1000 $t_g$. The equatorial outflow marked by the outward velocity streamline expands in time as the simulation proceeds. Shocks can be noticed in the velocity field structure.}
    \label{fig:vel_500rg}
\end{figure*}

The magnetic flux on the black hole horizon is quantified by integrating the radial component of the magnetic field over the horizon and by normalizing it to the inward mass flux,

\begin{equation}
    \phi_\mathrm{BH}(t) = \frac{1}{2\sqrt{\dot{M}}} {\int_{\theta} \int_{\phi} \mid{B^{r}(r_{H},t)}\mid dA_{\theta \phi}}.
    \label{eq:phi_BH}
\end{equation}

Figure \ref{fig:[b01_2D]_phiBH_t} shows the time-evolution of magnetic flux on the black hole horizon for the models with initial maximum $\beta = 0.1$. 
The initial flux is high as the initial uniform field strength is much higher than the gas pressure. As the accretion proceeds, this flux diminishes, but keeps values of $\phi_\mathrm{BH} \sim 30$--$45$ as normalized to the inward mass flux. Values in this range suggest that the accretion is magnetically arrested and further inflow of matter can occur through magnetic reconnections and interchange instabilities. 

The mass outflows are sustained in time as can be seen from the outflow rates plot (again for the models with $\beta = 0.1$) shown in Figure \ref{fig:[b01_2D]_mout_t}, computed at $10~r_g$. This is the case with all our models with the highly magnetized models showing slightly higher outflow rates when averaged over time, compared to the models in equipartition of gas to magnetic pressure, while being in the same order of magnitude. 

Figure \ref{fig:[b01_2D]_mdot_t_log} shows the inward mass accretion rate at the black hole horizon for the models with $\beta = 0.1$. We can notice some quasi-periodic fluctuations in the mass accretion rate as well as the magnetic flux on the horizon (in Figure \ref{fig:[b01_2D]_phiBH_t}) and these effects seem to depend on the black hole spin. So we can possibly attribute the fluctuating behaviour of the flux on the black hole horizon (and in turn the mass accretion rate) to the expulsion of magnetic field lines by a highly spinning black hole. From these plots, it is also evident that the model with highest black hole spin shows the highest mass outflow rate as well the highest mass accretion rate. 

From the models presented in Table \ref{tab:eq_outflow_models}, we note that the black hole spin has a clear influence on the rate of mass outflows. The outflows rates show an increasing pattern with the black hole spin parameter ranging from $a = 0.60$ to 0.99, systematically in all models with varying magnetic field strength. This can be attributed partly to the Meissner-like expulsion of magnetic field lines by the rotating black holes which increases with the spin of the black hole, which in turn reflects in the outflow rates. At the same time, the strength of magnetization does not show a clear influence on the rate of outflows. All the models varying from $\beta=0.1$ to 1 show similar outflow rates without a clear trend. 
On the other hand, the black hole spin shows a less pronounced effect on the inward mass accretion rate. The accretion rates reaches similar values for the dimensionless spin values of $a=0.60$ and 0.90 in all levels of magnetization. And it shows slightly higher values for the near to maximally spinning cases. But, the Schwarzchild cases show higher accretion rates than the Kerr black holes, except in the most highly magnetized case. The table also lists the time averaged values of the dimensionless flux of the magnetic field on the black hole horizon ($\phi_\mathrm{BH}$).

\begin{table}[tbh!]
\caption[Unit conversions]{Example for the conversion of geometric to cgs units. Here we adopt a black hole mass of $M = 6.2 \times 10^9 M_\odot$ considering the M87*}
\label{tab:unit_conv}
\centering
\begin{tabular}{l l l}
\hline
Physical quantity & Geometric units & cgs units \\
\hline
Length & $r_g = GM/c^2$ & $9.159 \times 10^{14}$ cm\\
Time & $t_g = GM/c^3$ & $3.055 \times 10^4$ s\\
\hline
\end{tabular}
\end{table}


\begin{figure}[tbh!]
    \centering
    \includegraphics[width=0.47\textwidth]{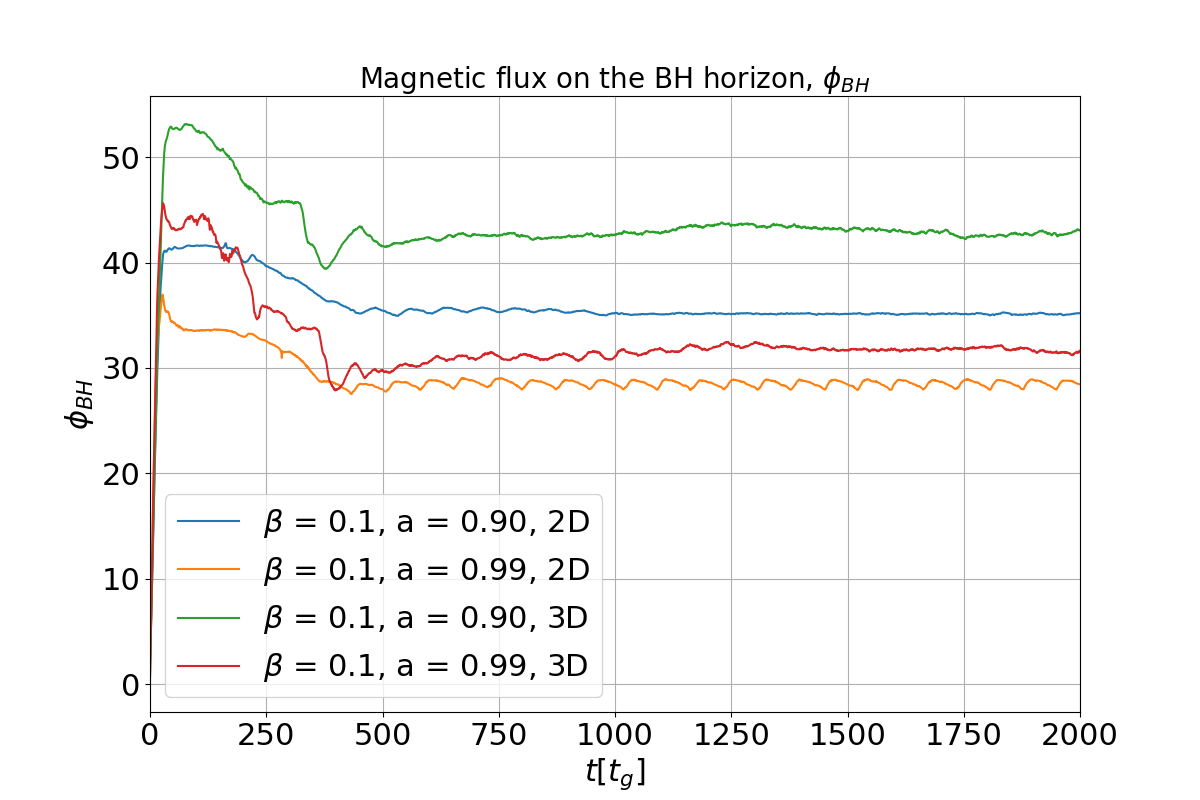}
    \caption[$\phi_\mathrm{BH}$ with time for the 3D models with $\beta = 0.1$]{The evolution of magnetic flux on the black hole horizon with time for  the 3D models with $\beta = 0.1$ and with different spin values from time $t = 1000~t_g$ (values from 2D models are for comparison).}
    \label{fig:[b01_3D]_phiBH_t}
\end{figure}

\begin{figure}[tbh!]
    \centering
    \includegraphics[width=0.47\textwidth]{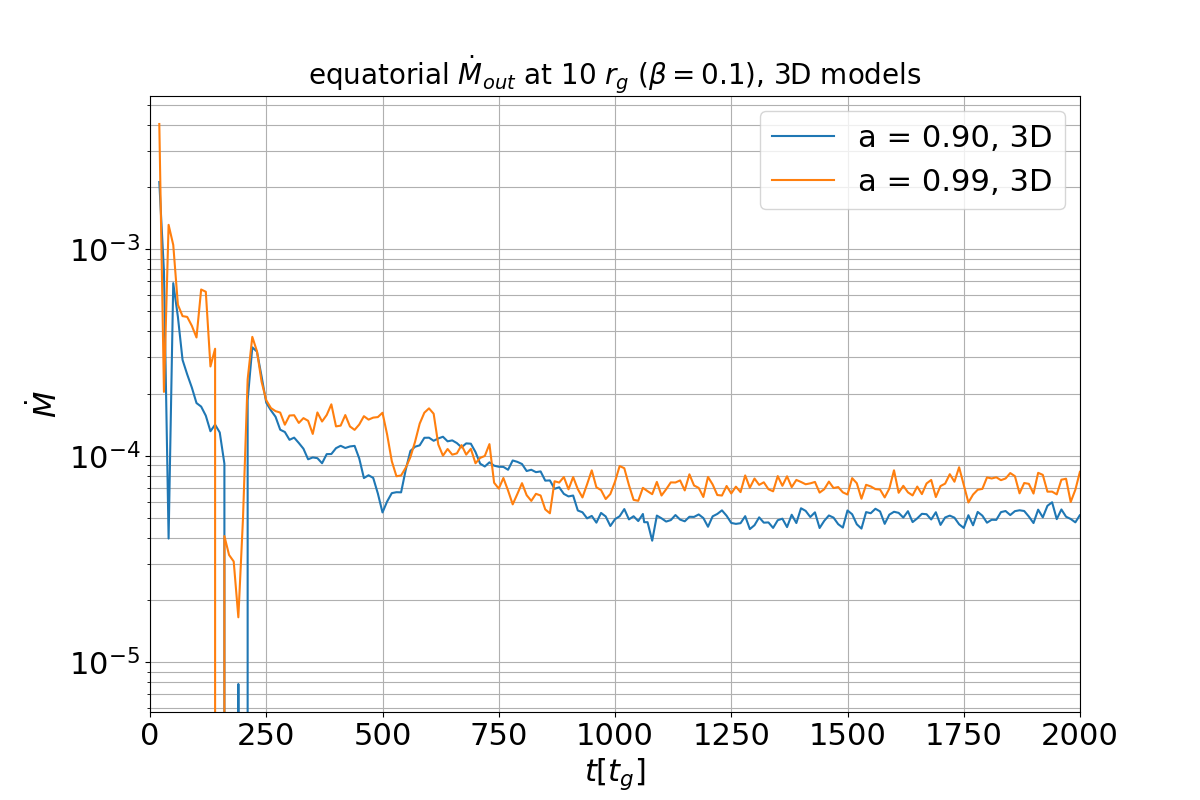}
    \caption[Mass outflow rate with time for the 3D models with $\beta = 0.1$]{The equatorial mass outflow rate with time at 10 $r_g$, after the magnetic field is turned on, for  the 3D models with $\beta = 0.1$ and different spin values} 
    \label{fig:[b01_3D]_mout_t}
\end{figure}

\begin{figure}[tbh!]
    \centering
    \includegraphics[width=0.47\textwidth]{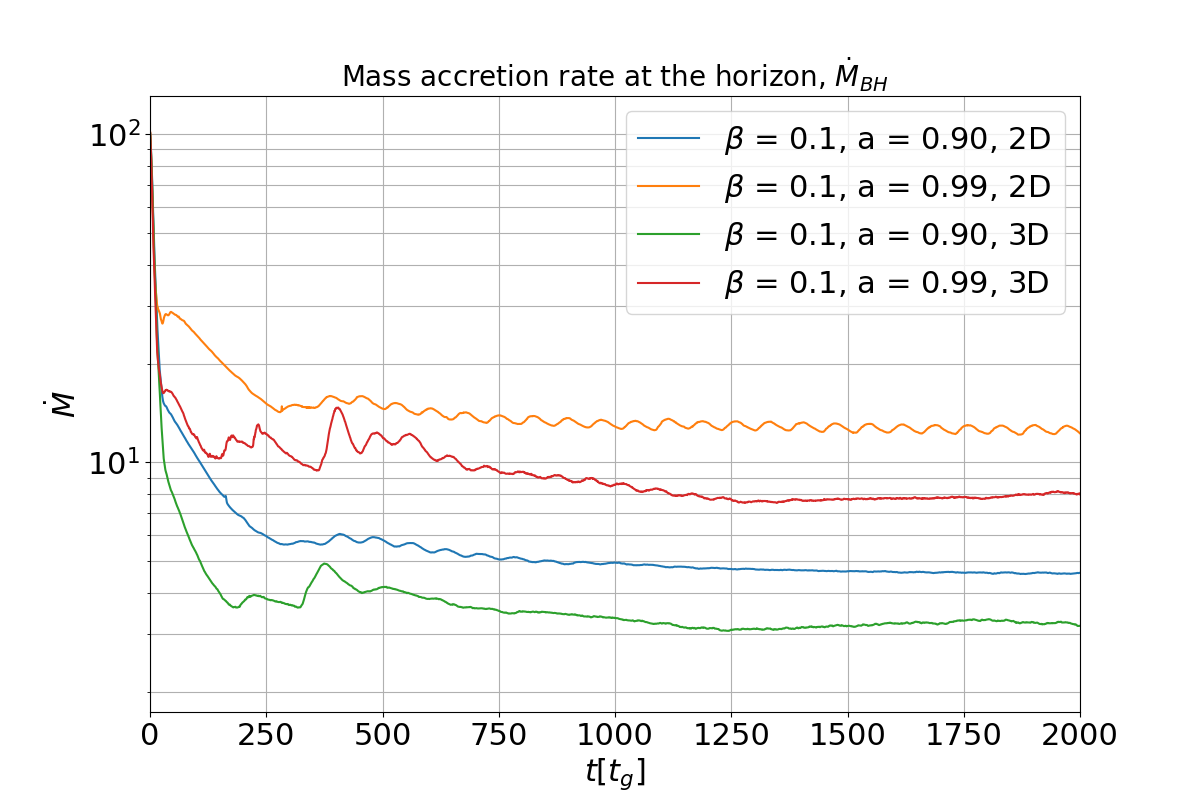}
    \caption[Mass accretion rate with time for the 3D models with $\beta = 0.1$]{The evolution of inward mass accretion rate at the black hole horizon with time, after the magnetic field is turned on, for the 3D models with $\beta = 0.1$ and different spin values (values from 2D models are for comparison).}
    \label{fig:[b01_3D]_mdot_t_log}
\end{figure}

In the last column of Table \ref{tab:eq_outflow_models}, we give the estimated mass outflow rate for a realistic physical system considering its known parameters. For this, we converted our results from the code units into physical units as follows. Since we use a unit convention of $G = c = M = 1$ in the code, the length and time units for the simulations results are given by $L_\mathrm{unit} = GM/c^2$ and $T_\mathrm{unit} = GM/c^3$ respectively. So the length and time values can be converted to physical units by using the relevant value of the black hole mass. Table \ref{tab:unit_conv} lists the conversion of geometric to physical units considering the black hole mass for M87*. Now, the density unit of the plasma around the black hole is related to the length unit by $\rho_\mathrm{unit} = M_\mathrm{scale}/L_\mathrm{unit}^3$ so that $M_\mathrm{scale}$ will depend on the environment under consideration. This density scaling for the AGN accreting environments is rather arbitrary (than for GRBs) as the spatial extension of the plasma is large and the density drops by many orders of magnitude from the black hole to the broad line region \citep{Czerny2016ApJ...832...15C}. In Table \ref{tab:eq_outflow_models}, we give the outflow rates from our model for the M87 central engine assuming a black hole mass of $6.2 \times 10^9 M_\odot$ and a density scaling of $\rho_\mathrm{unit} = 8.85 \times 10^{-18} \mathrm {g~cm}^{-3}$ \citep{Janiuk_James_2022A&A}. 

\begin{figure*}[tbh!]
    \centering
    
    \includegraphics[width=0.33\textwidth]{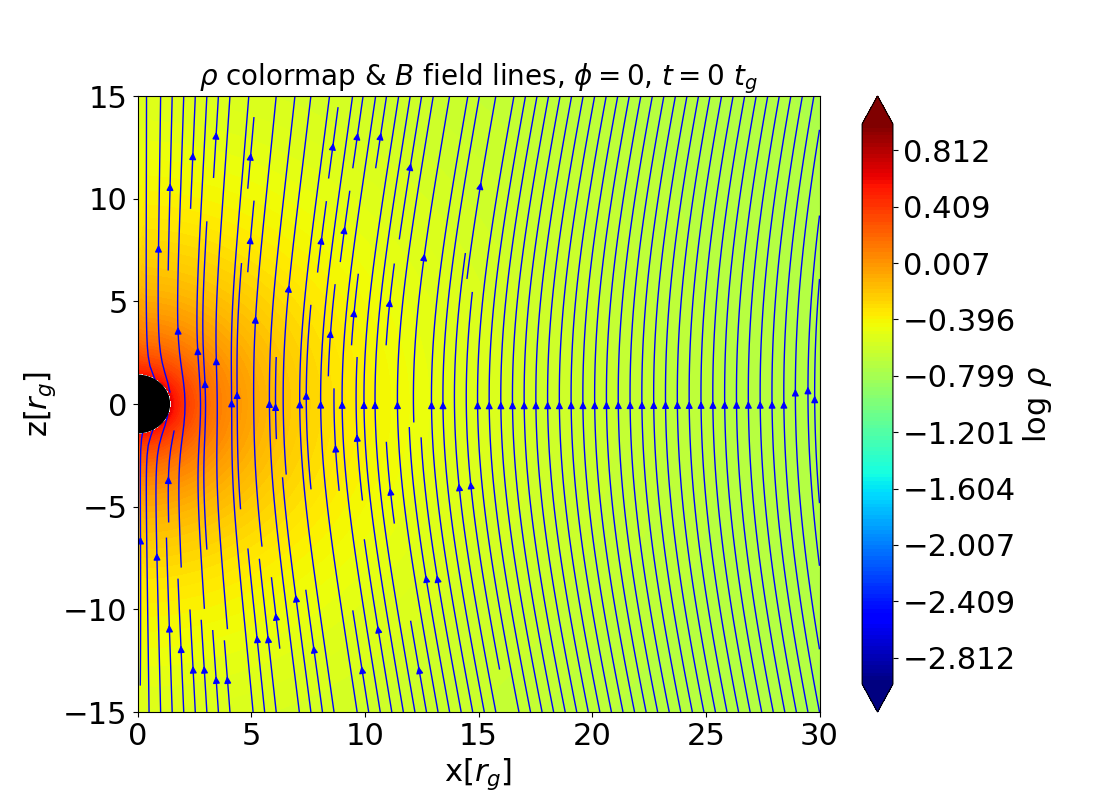}
    \includegraphics[width=0.33\textwidth]{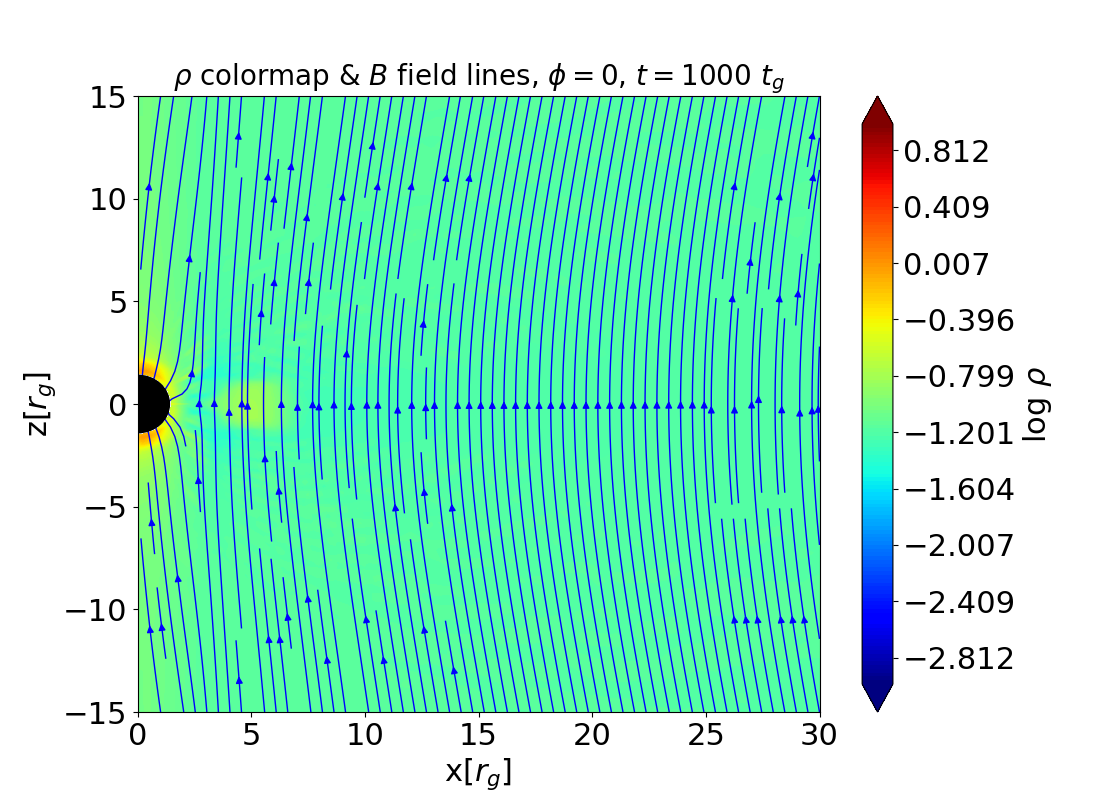}
    \includegraphics[width=0.33\textwidth]{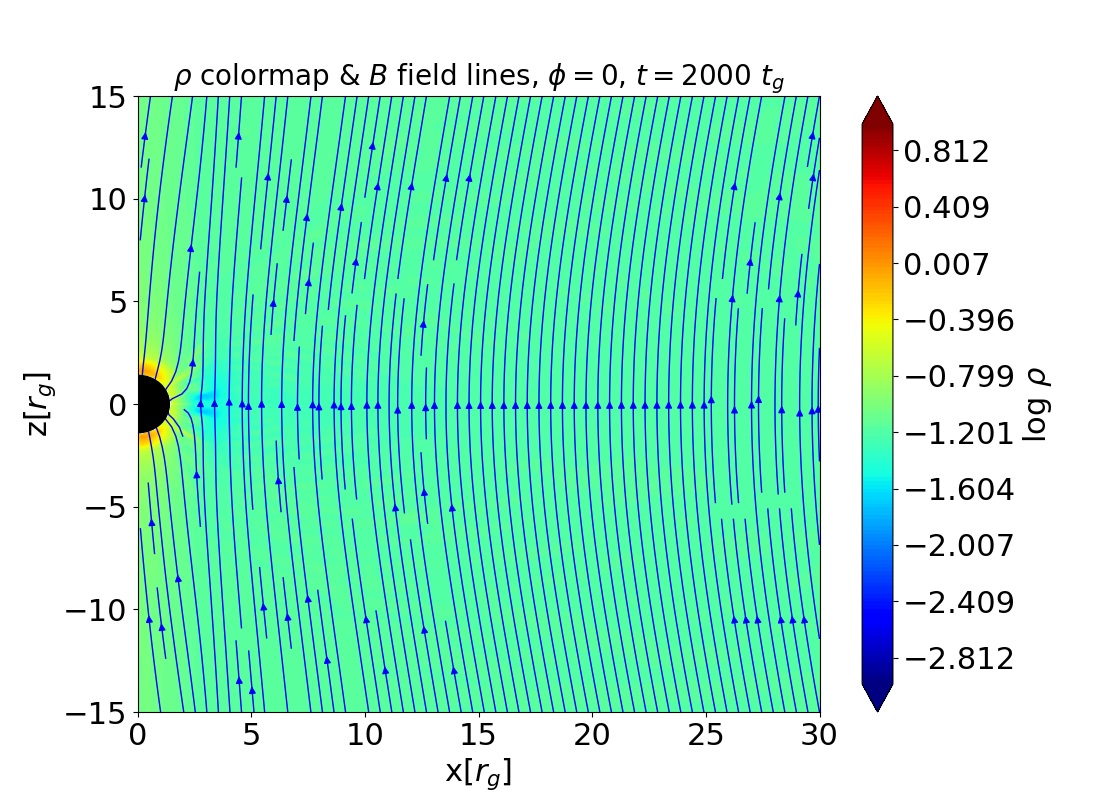}
    
    \includegraphics[width=0.33\textwidth]{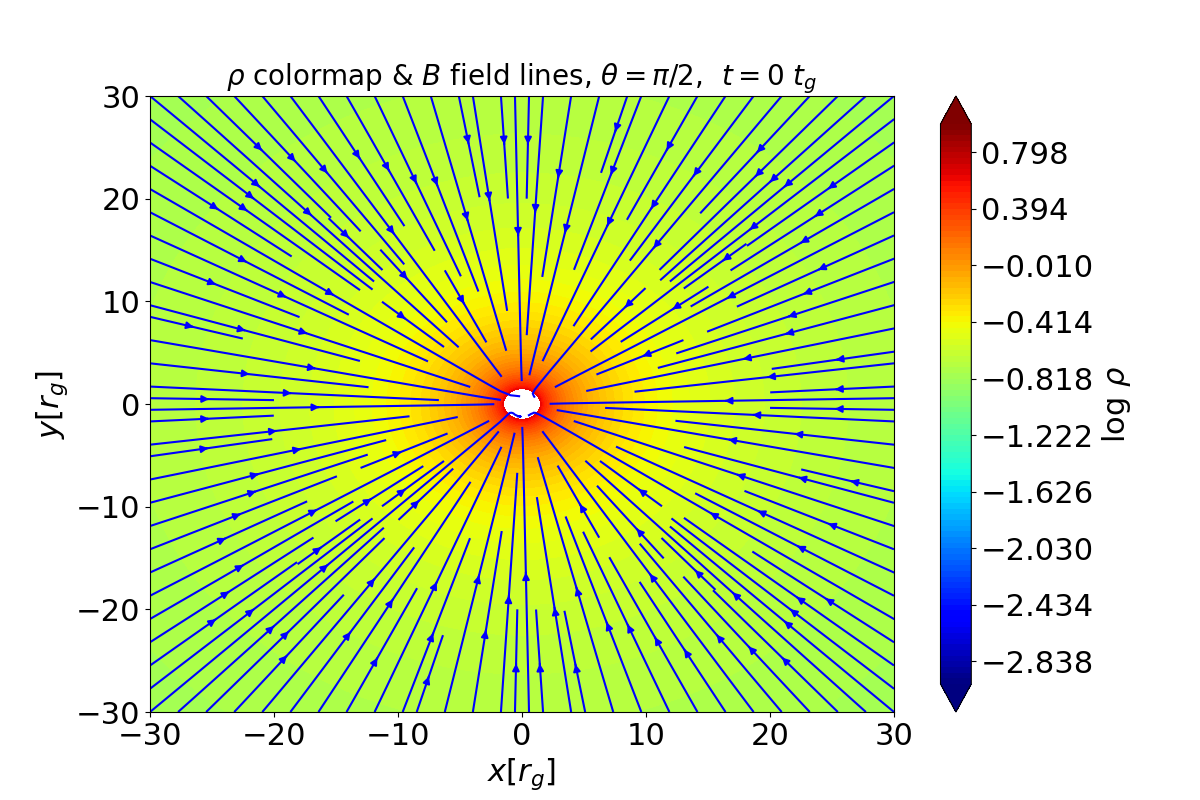}
    \includegraphics[width=0.33\textwidth]{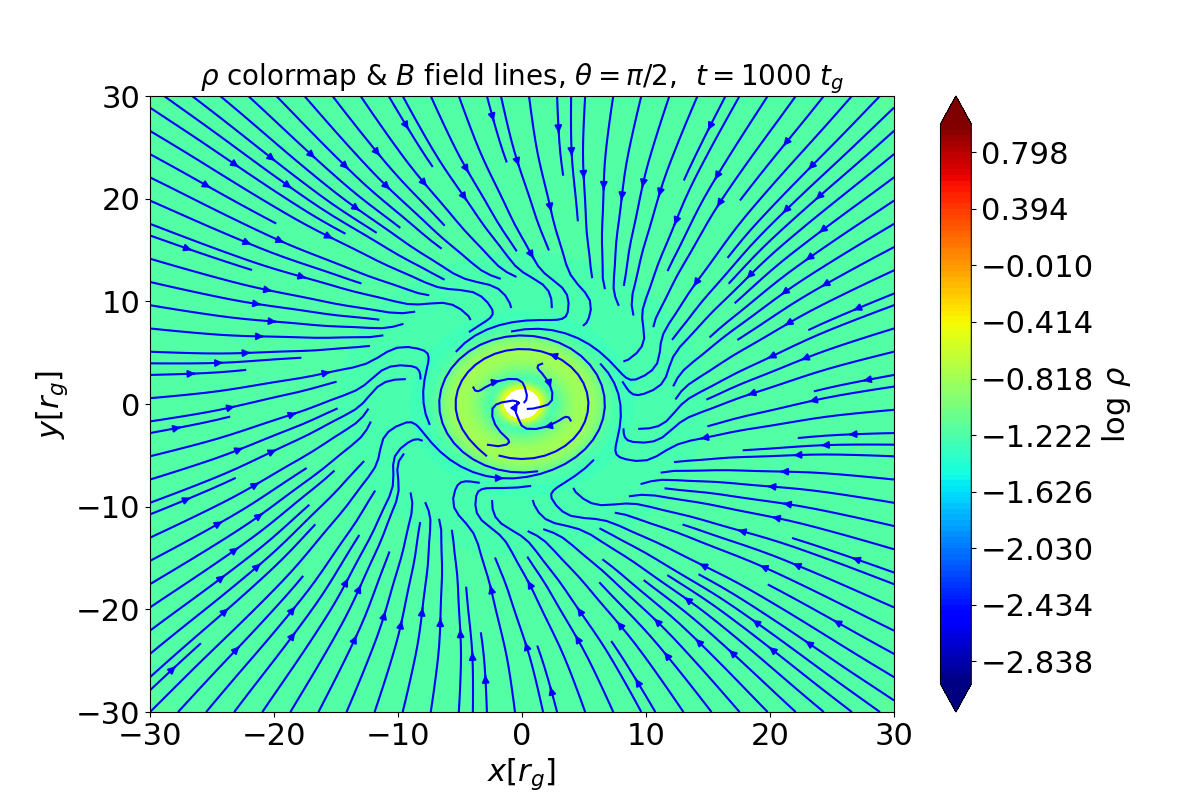}
    \includegraphics[width=0.33\textwidth]{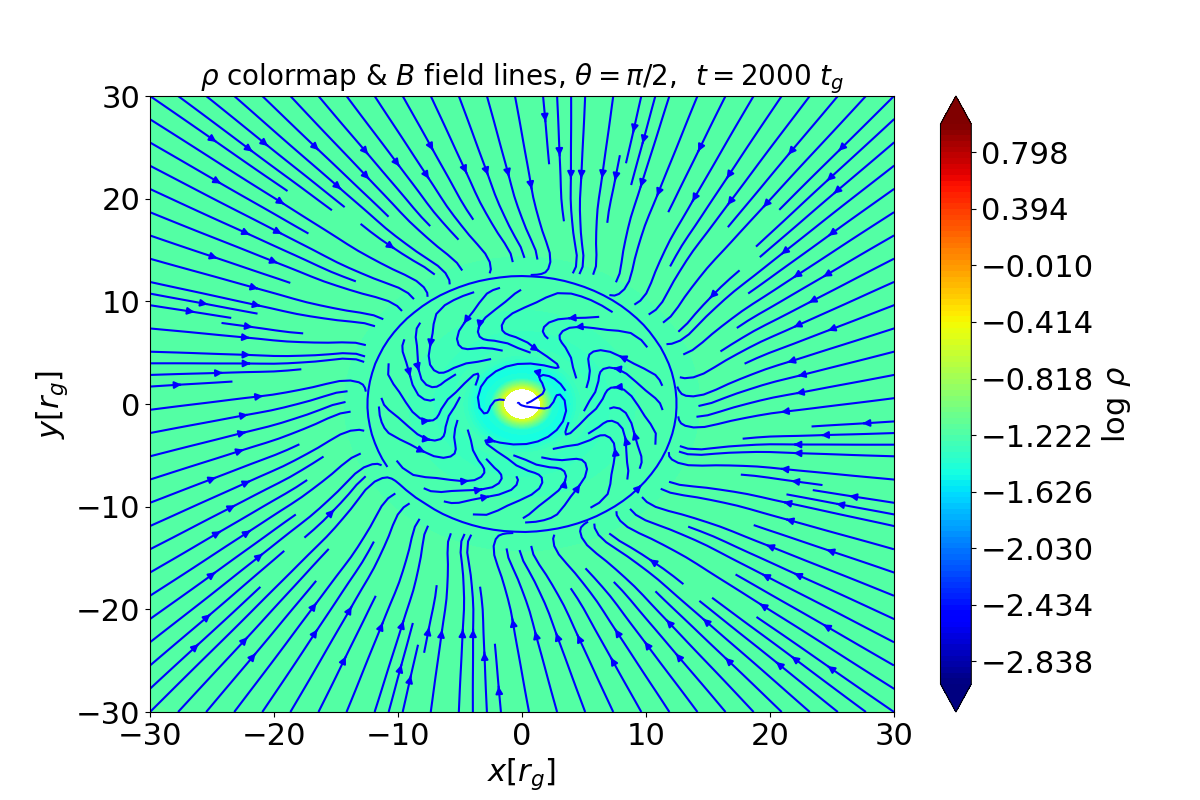}

    \includegraphics[width=0.33\textwidth]{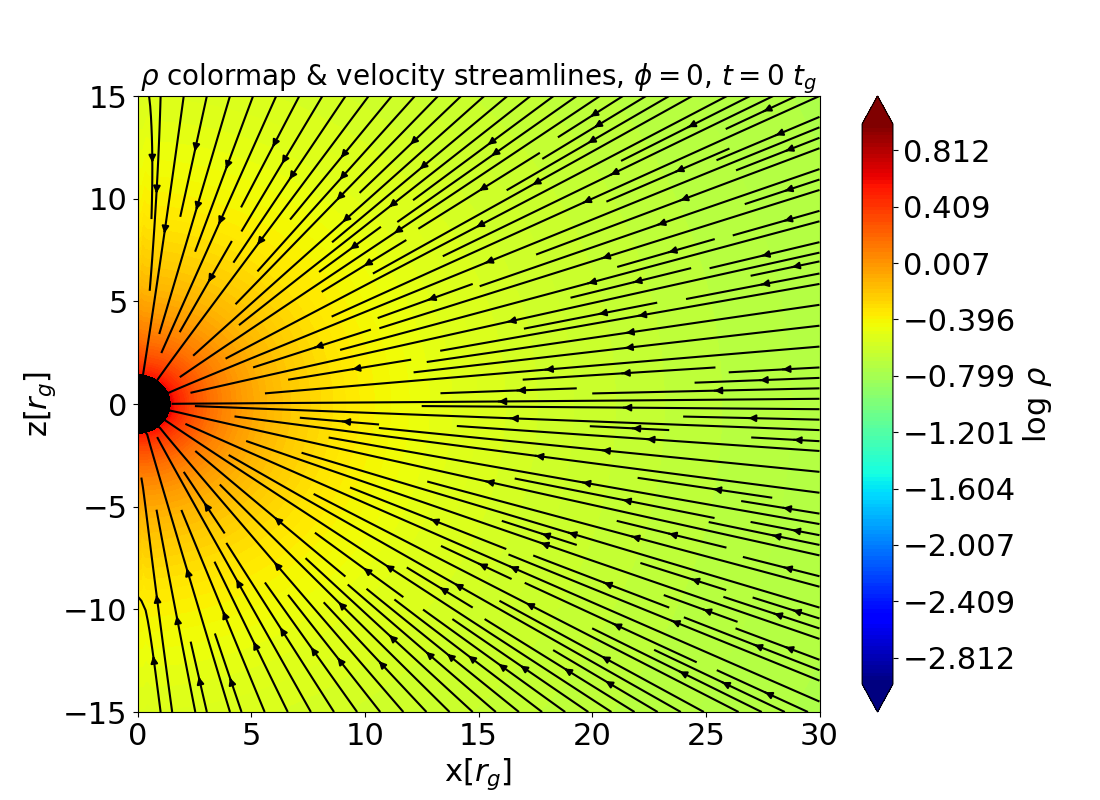}
    \includegraphics[width=0.33\textwidth]{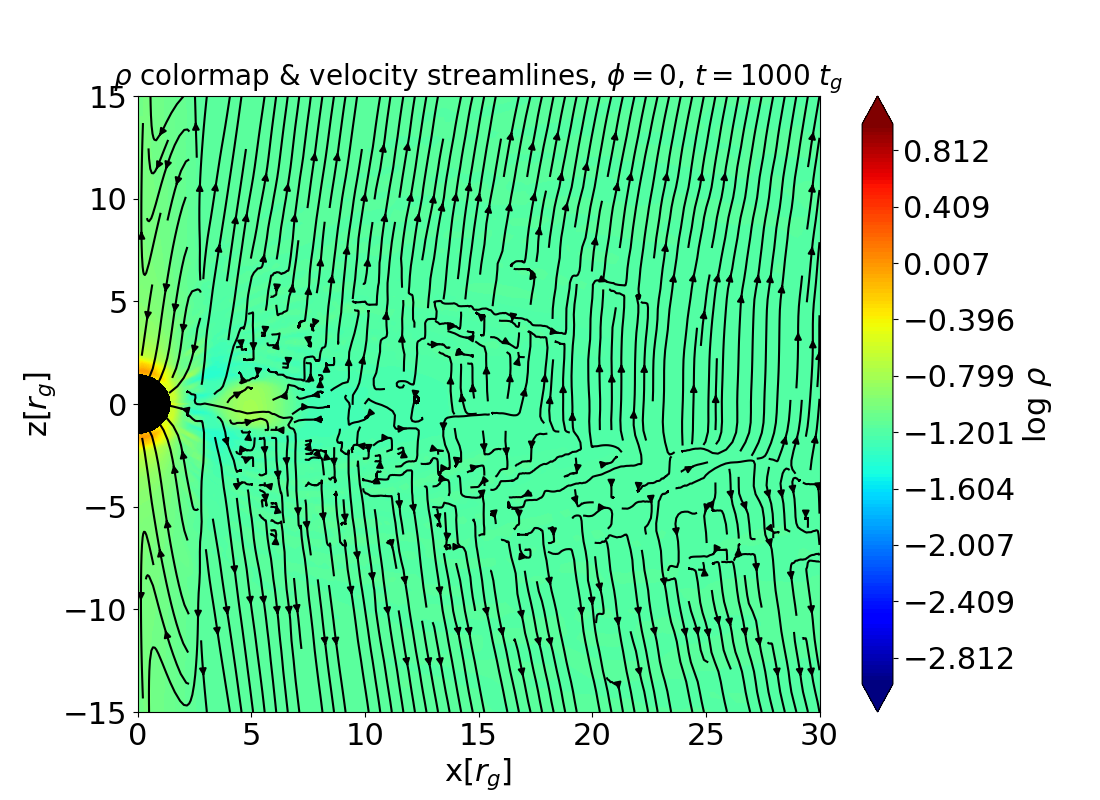}
    \includegraphics[width=0.33\textwidth]{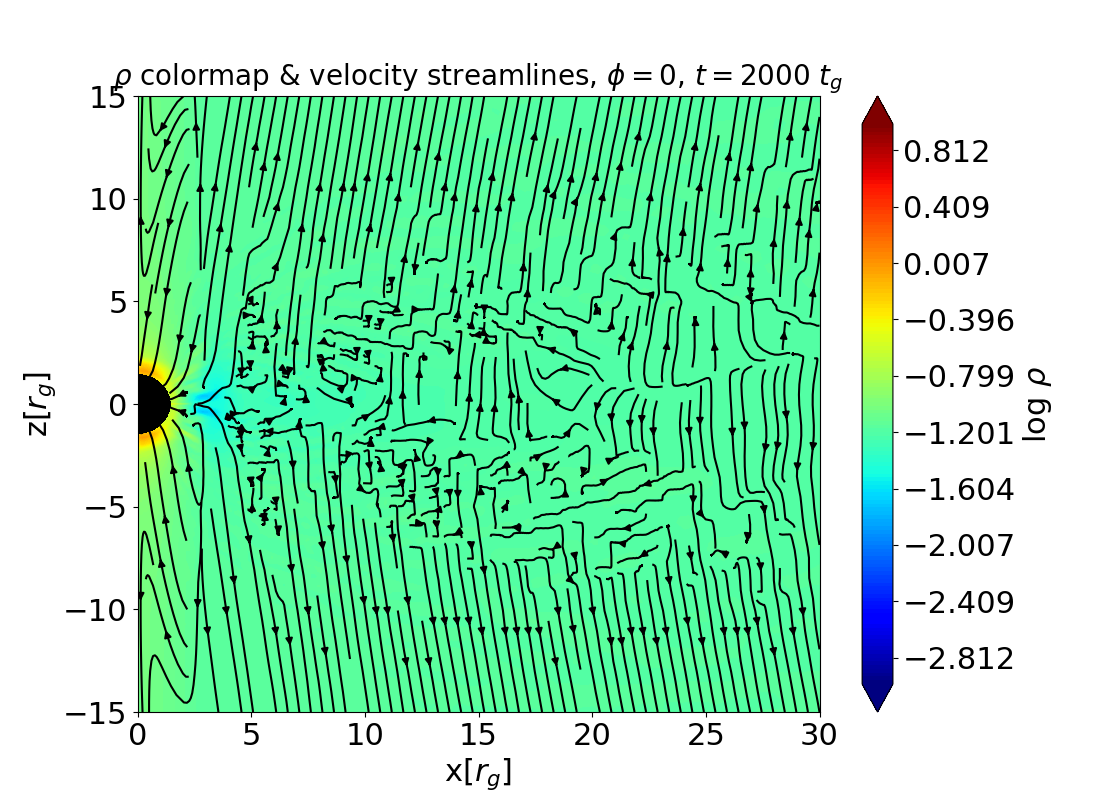}

    \includegraphics[width=0.33\textwidth]{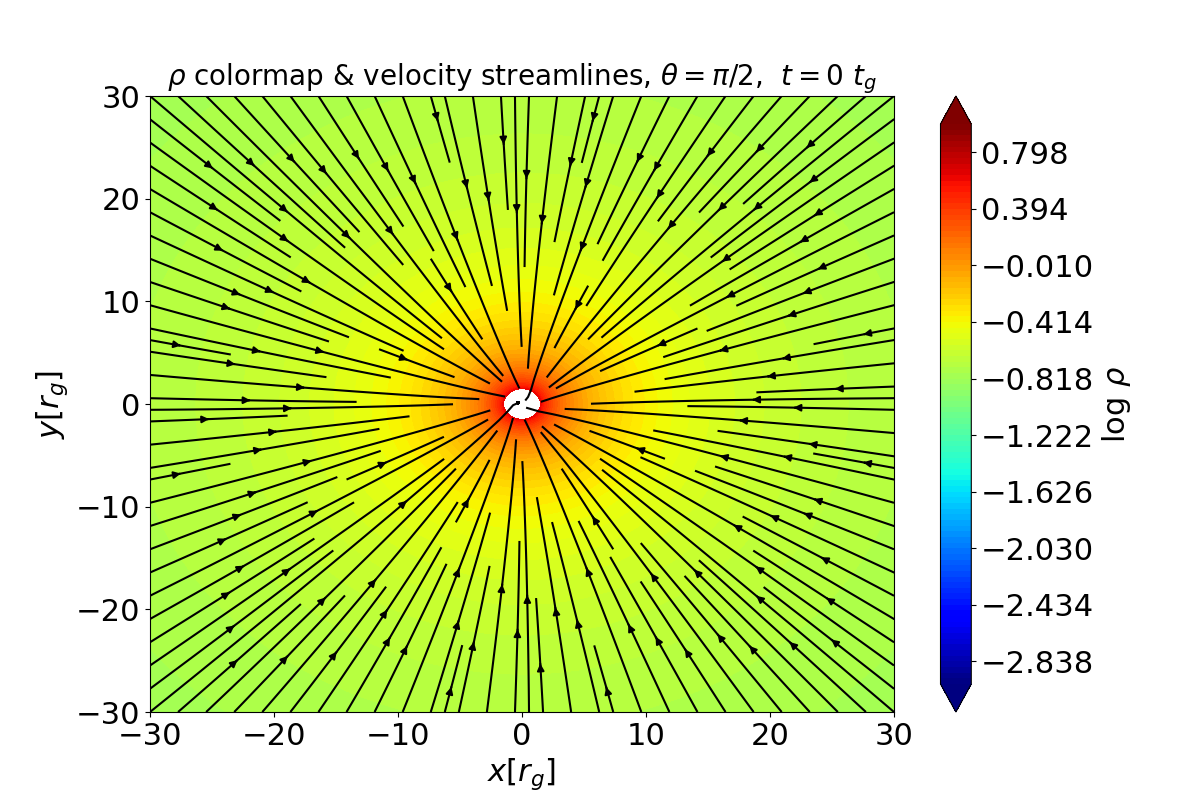}
    \includegraphics[width=0.33\textwidth]{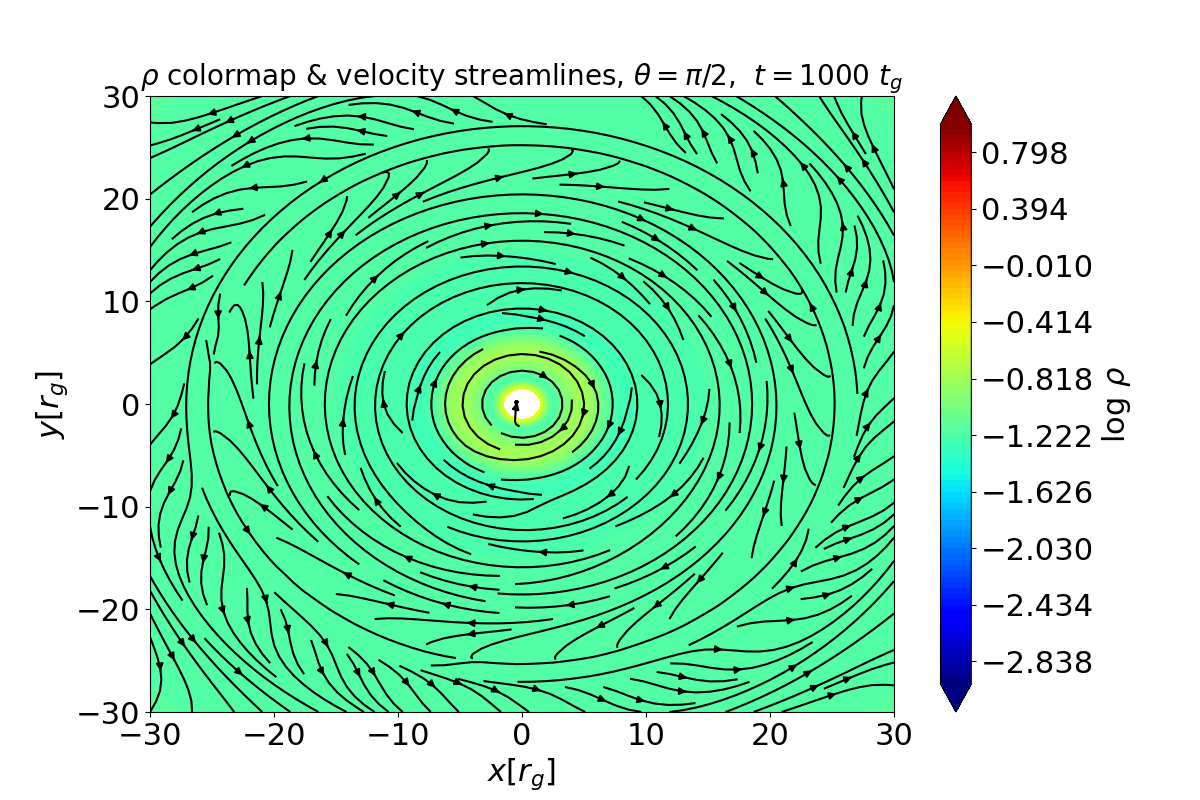}
    \includegraphics[width=0.33\textwidth]{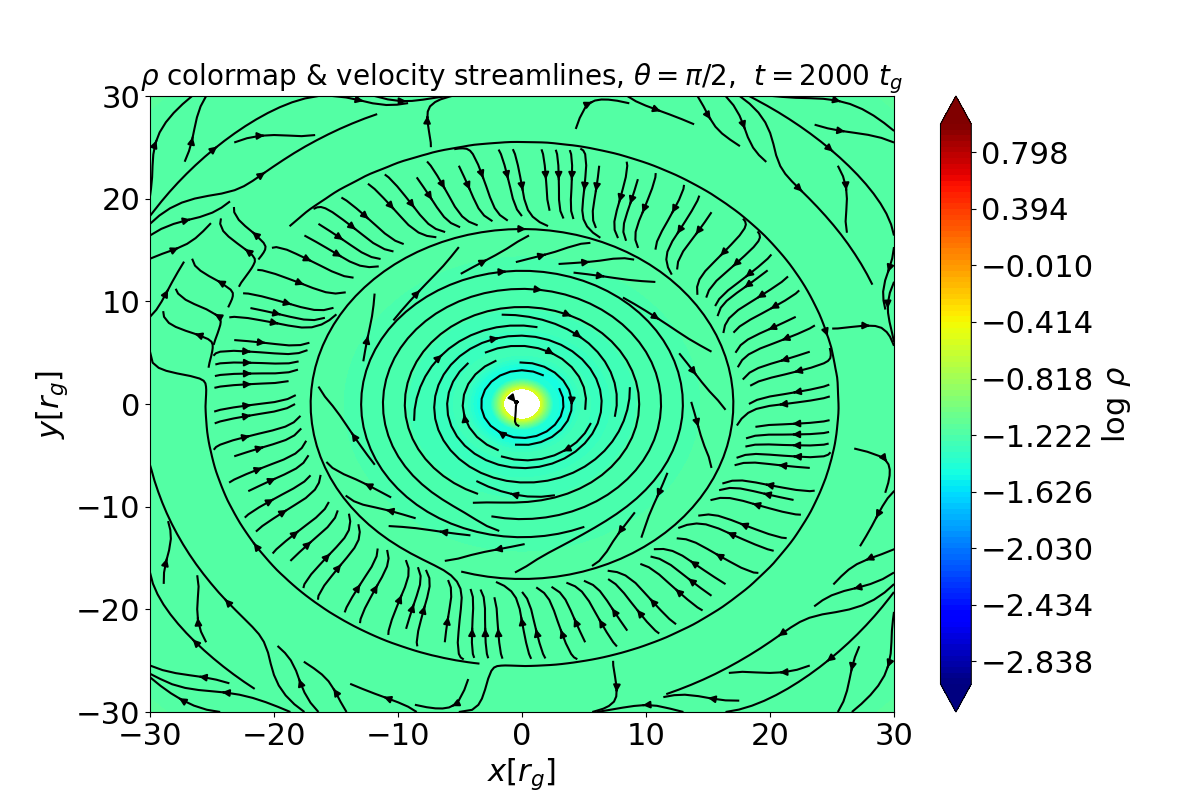}
    \caption{Time evolution of the aligned magnetic field 3D model: Selection of exemplary plots exhibits density, magnetic field (two top panels), and velocity (two bottom panels) evolution for the b01.a90.3D model. 
        The equatorial plane $(x,y)$ (view along the rotation axis) and the poloidal section $(x,z)$ (view perpendicular to the rotation axis) are shown. The left-most panels show the initial configuration with the magnetic field tuned on. The the middle and the right panels show the later evolved states. The first row depicts poloidal slices at $\phi = 0^\circ$, the second row shows equatorial slices ($\theta = \pi/2$). The third and fourth rows show similar slices for velocities.  
    } 
    \label{fig:[b01_a90_3D]init_evolvd_density_B_pol}
\end{figure*}

We notice the discontinuities in velocity while the outflow is expanding, which also means that the eruptions are discontinuous. But with time, the outflows continue towards the outer boundary of the computational domain.
The expansion of developed outflows to larger distances in one of our models (b01.a99.2D) is demonstrated in Figure \ref{fig:vel_500rg} with velocity streamlines. It should be noted that the flow remains symmetric to the equator, on average over a longer period of time, even though some asymmetric behaviour with respect to the equator can be observed at certain time instances.

\begin{figure*}[tbh!]
    \centering
    \includegraphics[width=0.33\textwidth]{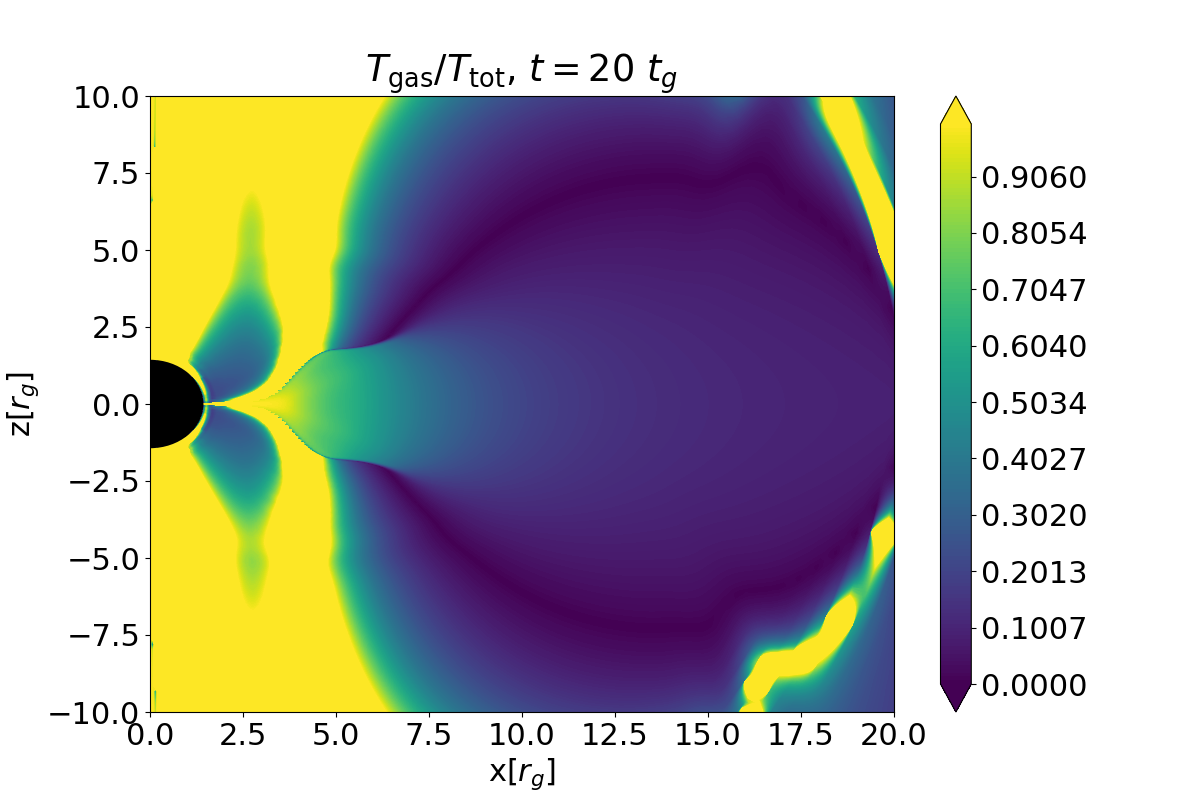}
    \includegraphics[width=0.33\textwidth]{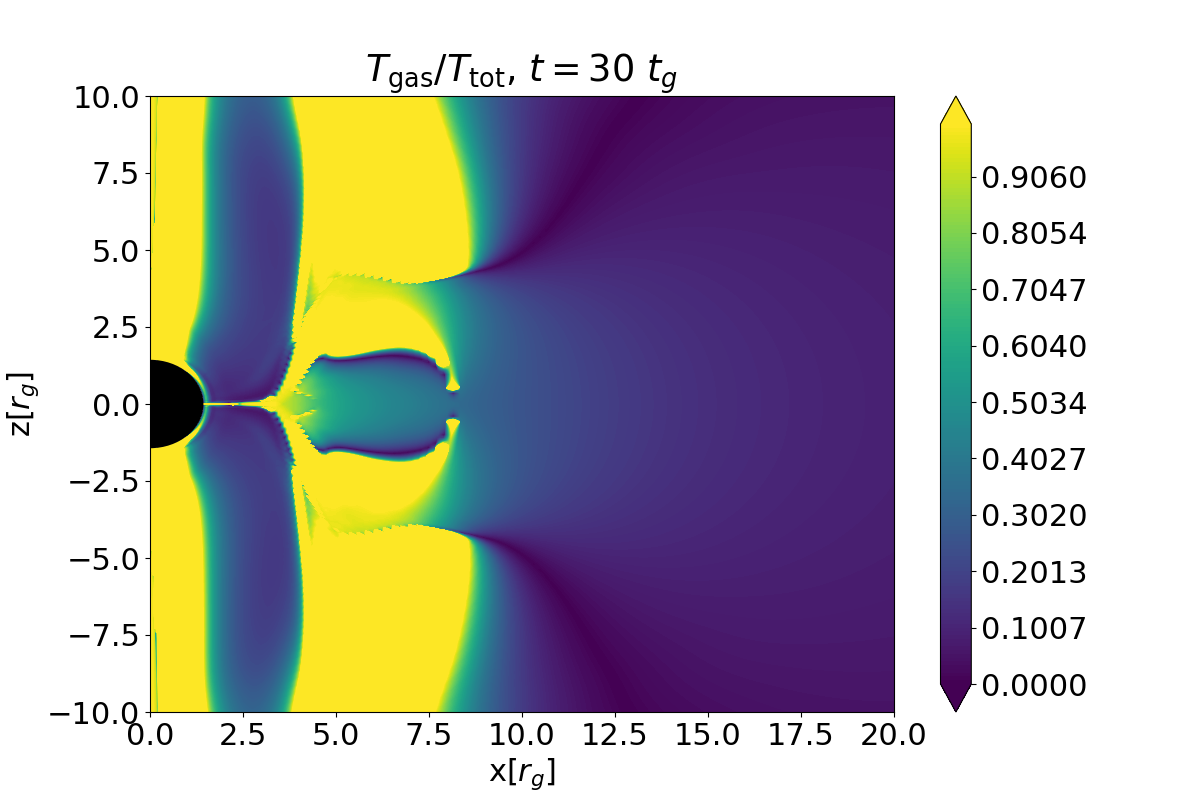}
    \includegraphics[width=0.33\textwidth]{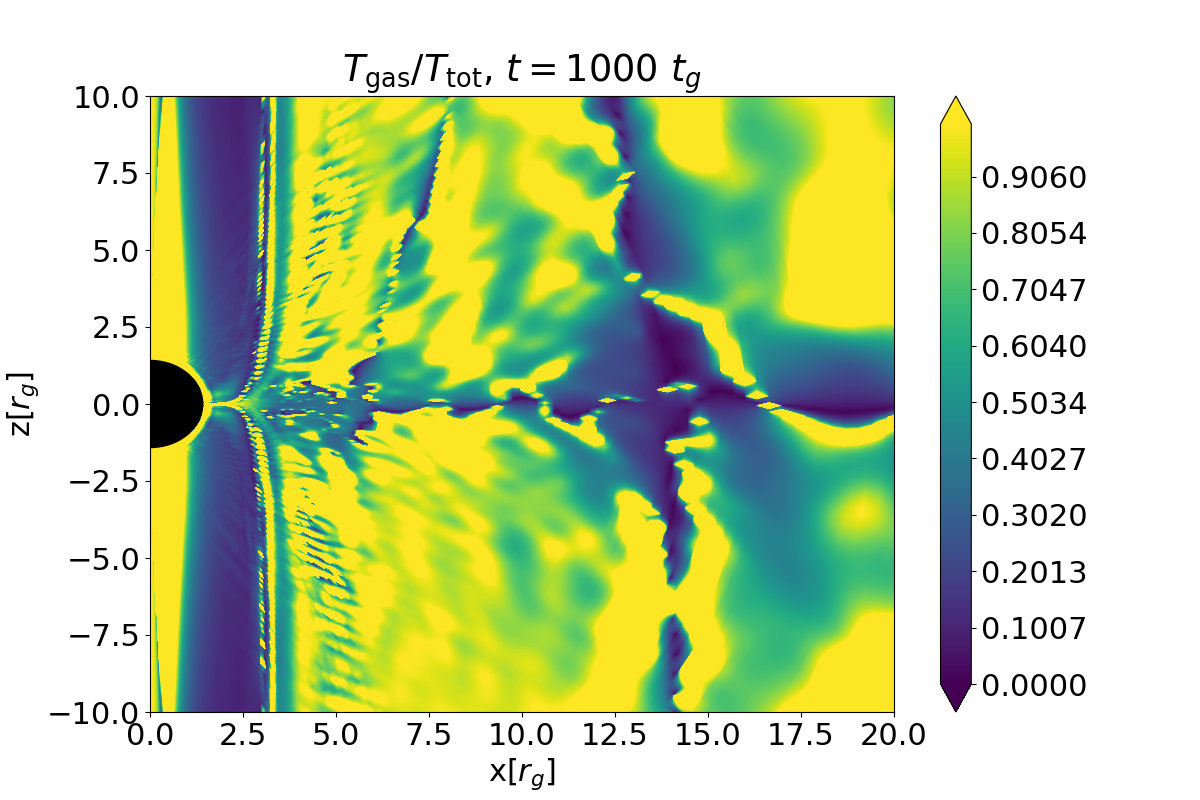}
    \includegraphics[width=0.33\textwidth]{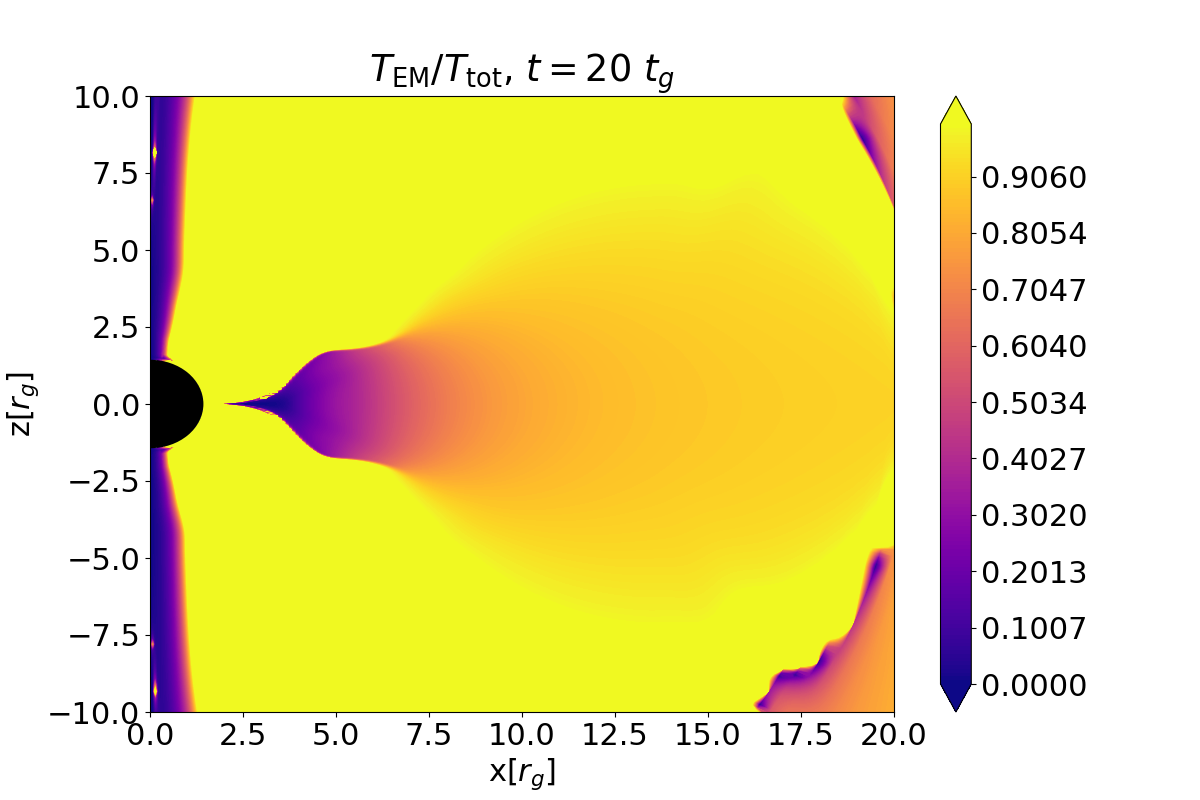}
    \includegraphics[width=0.33\textwidth]{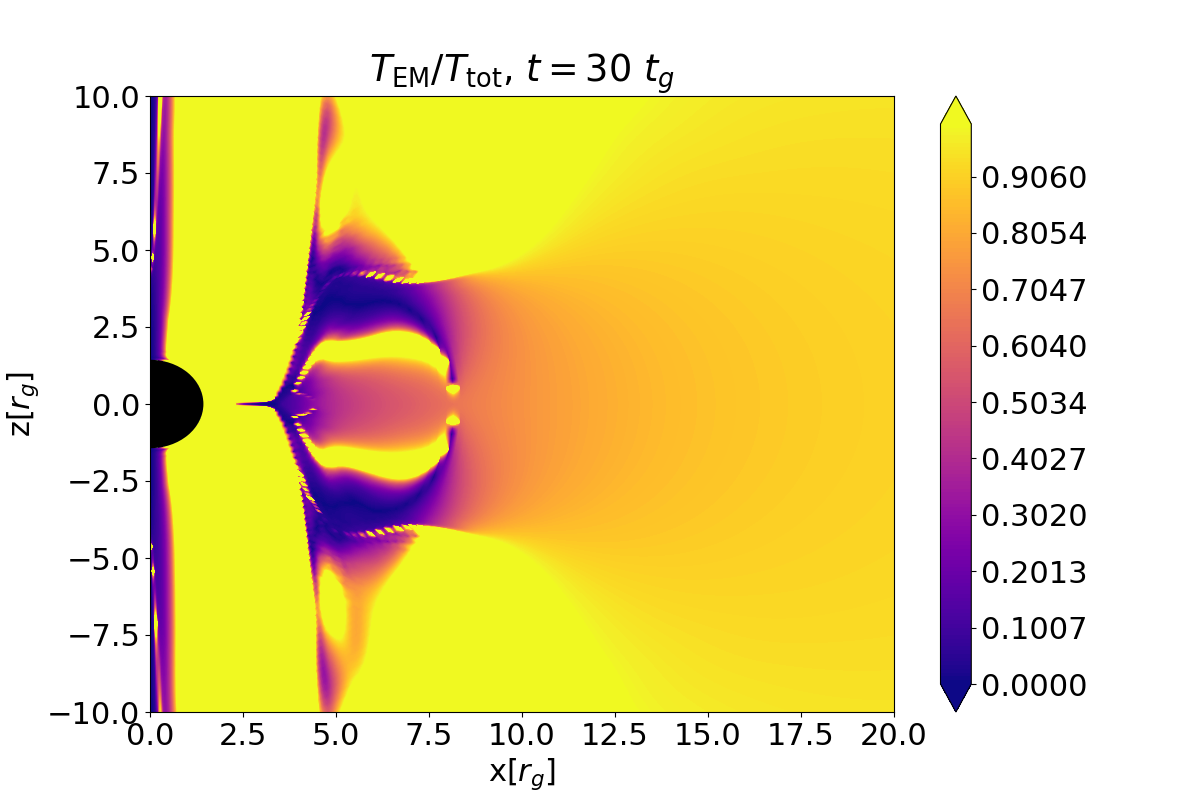}
    \includegraphics[width=0.33\textwidth]{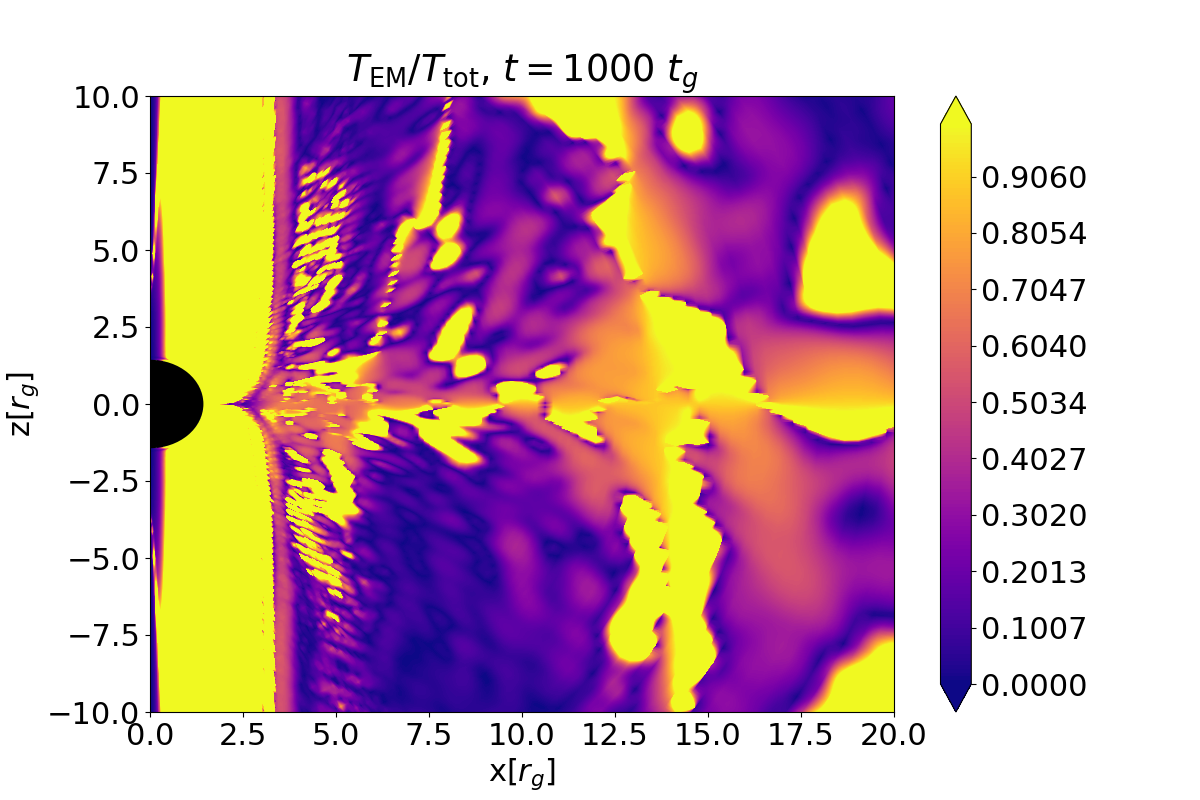}
    \caption{The energy composition of the developed outflows in a representative model b01.a90.2D. Plots show the ratio of the matter (top row) and electromagnetic energy (bottom row) to the total energy, at chosen time instances: 20, 30 and 1000 $t_g$. While the system is relatively organized at the beginning of the evolution, turbulent behaviour gradually prevails in the course of time. Large-scale structures are still seen in the equatorial plane and along the axis, although a well-defined jet does not develop in this simulation. The value in the color-scale represents a fraction ranging from 0 to 1.} 
    \label{fig:Energy_MA_EM}
\end{figure*}

We also investigated two models in 3D, with the same initial configuration (for $a=0.90$ and 0.99) for the highly magnetized case ($\beta=0.1$) to more correctly account for the evolution of the system. 
In Figure \ref{fig:[b01_3D]_phiBH_t}, we plot the magnetic flux on the black hole horizon in the case of the 3D models. The values from the 2D models with the same parameters are included for comparison. From this we notice that the flux is slightly lower for the 3D models with the same parameters, but they both are in the same order of magnitude as well as in the magnetically arrested accretion state. We can also notice that there are less fluctuations in these values for the 3D models. Figure \ref{fig:[b01_3D]_mout_t} shows the azimuthally averaged equatorial outflow rate with time, computed at 10 $r_g$, for the 3D models. They are in the same order of magnitude as in the 2D ones, and the time averaged values are given in Table \ref{tab:eq_outflow_models}. From the values given the table, we notice that the outflow rate is systematically higher in both the 3D models compared to their 2D counterparts. The possibility for outflows along different azimuthal directions might be a reason for this. Thus our 2D models tend to underestimate the outflow rates. Finally, Figure \ref{fig:[b01_3D]_mdot_t_log} shows the inward mass accretion rate at the black hole horizon for the 3D models. We can notice the quasi-periodic fluctuations in the accretion rate are more smoothed out in the 3D models similar to the fluctuations in the magnetic flux.

Figure \ref{fig:[b01_a90_3D]init_evolvd_density_B_pol} show the initial and the evolved density states from the 3D models. The two top rows show the magnetic field liens for these models plotted on top of the density contours (on poloidal and equatorial slices), while the two bottom rows show the velocity streamlines at similar time instances of the evolution. As we have noticed in the 2D models, here also we notice equatorial outflows developed due the magnetic reconnection events in the equatorial region in the nearest vicinity of the black hole. By analyzing the equatorial slices, we notice that the outflows keep expanding in time. 

To better understand the energy composition of the developed outflows, 
we estimated the radial component ($T^r_t$) of the matter and electromagnetic parts of the stress-energy tensor, given by equations \ref{eqn:T_MA} and \ref{eqn:T_EMb}, for one of our representative model b01.a90.2D.
Figure \ref{fig:Energy_MA_EM} shows the ratio of the matter and electromagnetic parts of the energy to the total energy. The plots illustrate the case of a matter-dominated equatorial outflow in the very initial moments of the simulation (left and middle plots). This can be attributed to plasma being pushed out along the equatorial direction, as the (numerical) reconnection of magnetic field lines begins. At an evolved stage ($1000~t_g$, the right plot), the flow is mainly dominated by the electromagnetic contribution due to the presence of the persistent current sheet and the continued magnetic reconnection in the equatorial region and the outflows driven by it. The plot on the right also exhibits a higher electromagnetic energy content along the black hole rotation axis that is caused by accumulation of magnetic flux near the black hole horizon. 


\subsection{Models with magnetic field misaligned with respect to the BH rotation axis}

\begin{table*}[tbh!]

	\centering
	\caption[Summary of misaligned field models investigated]{Summary of two representative models with the magnetic field inclined with respect to the BH rotation axis. }
	\label{tab:incl_models}
	\begin{tabular}{lccccccc} 
		\hline
		Model & Angle of & $\beta_{min}$ & $a$ & Final & $\langle\phi_{\rm BH}\rangle_t$ & $\langle\dot M_{{\rm out,10r_{\it g}}}\rangle_t$ & $\langle\dot M_{{\rm out,10r_{\it g}}}\rangle_t$\\
            & inclination &   &  & $\dot M_{\rm in,H}$ &  & (code units) & ($M_{\odot}$ yr$^{-1}$) \\
		\hline

        incl45.bmin0003.a90 & $45^\circ$ & 0.0003 & 0.90 & 1.53 & 36.09 & $2.18 \times 10^{-4}$ & $7.69 \times 10^{-7}$ \\
        incl45.bmin0003.a99 & $45^\circ$ & 0.0003 & 0.99 & 2.68 & 29.12 & $3.21 \times 10^{-5}$ & $1.13 \times 10^{-7}$\\


		\hline
	\end{tabular}
 \tablefoot{The models are parameterized by the initial minimum plasma $\beta$ in the domain and the dimensionless Kerr parameter $a$. The time-averaged magnetic flux on the horizon and the mass outflow rate are computed 
 from time $1500~t_g$ to $2000~t_g$, after the magnetic field is turned on. The final mass accretion rate is given at $2000~t_g$ after the magnetic field has been turned on. The outflow rate in physical units is estimated by considering the M87 central engine mass, as for the earlier models. 

 }
\end{table*}

\begin{figure*}[tbh!]
    \centering
    
    \includegraphics[width=0.33\textwidth]{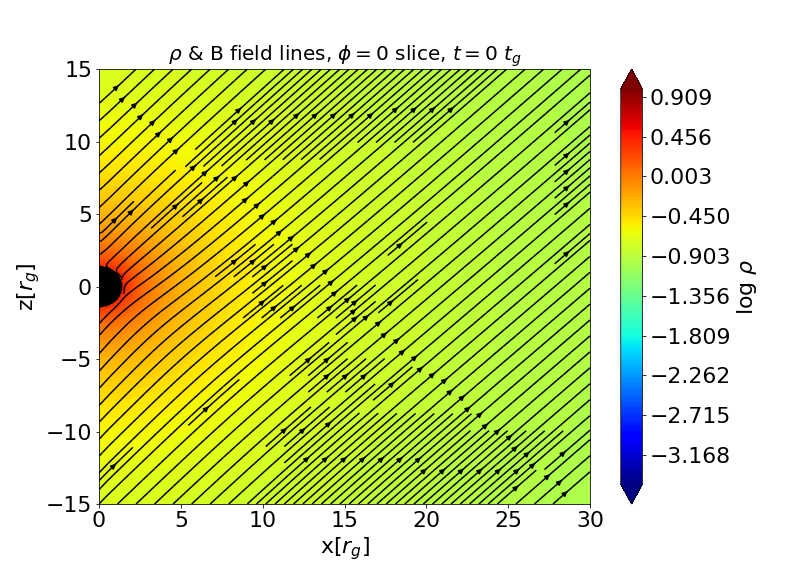}
    \includegraphics[width=0.33\textwidth]{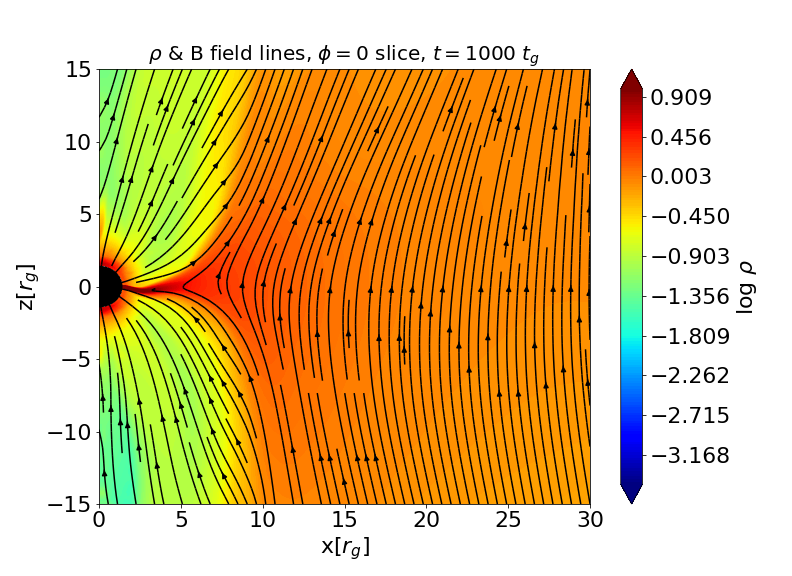}
    \includegraphics[width=0.33\textwidth]{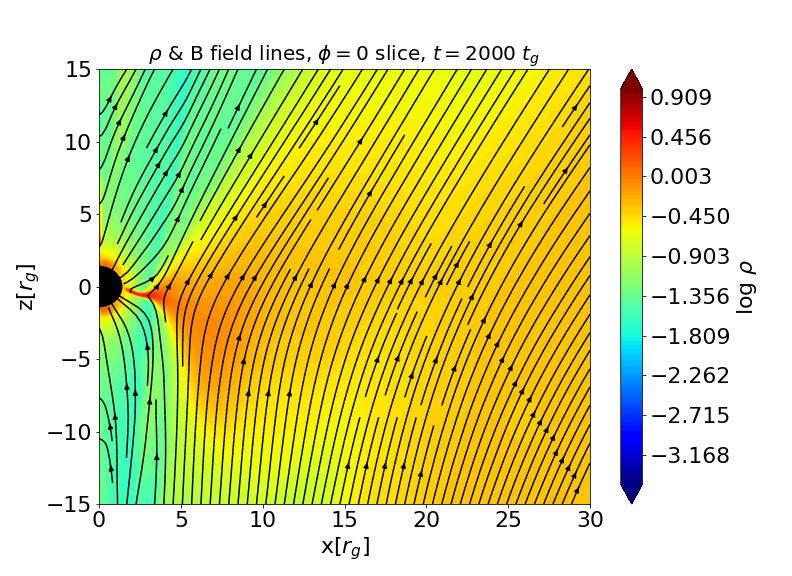}
    
    \includegraphics[width=0.33\textwidth]{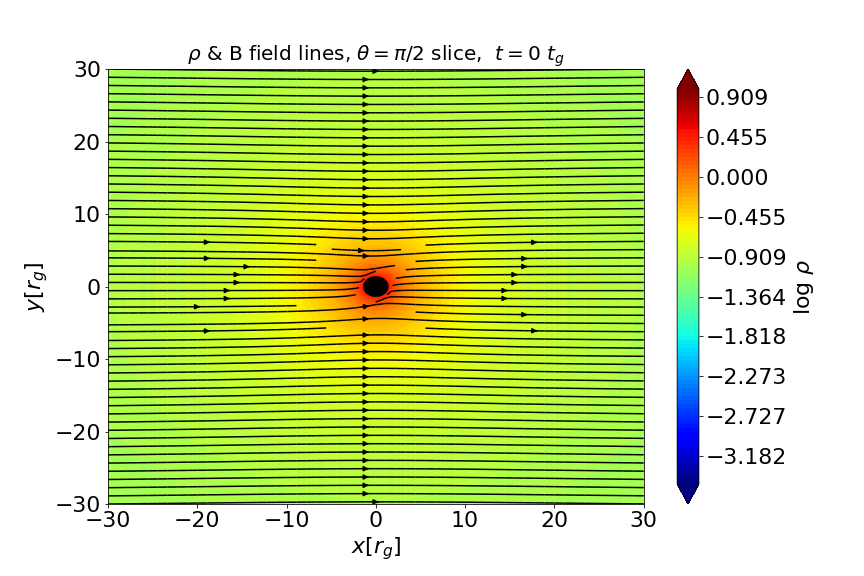}
    \includegraphics[width=0.33\textwidth]{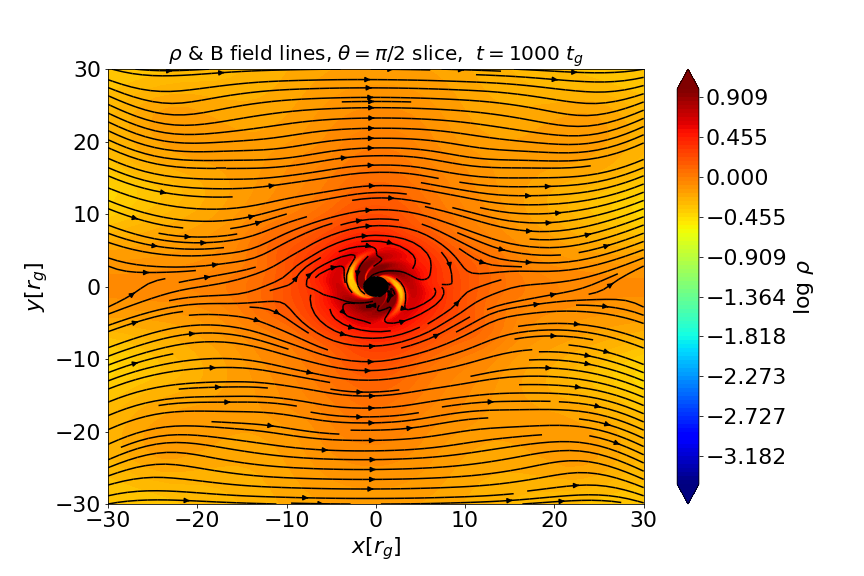}
    \includegraphics[width=0.33\textwidth]{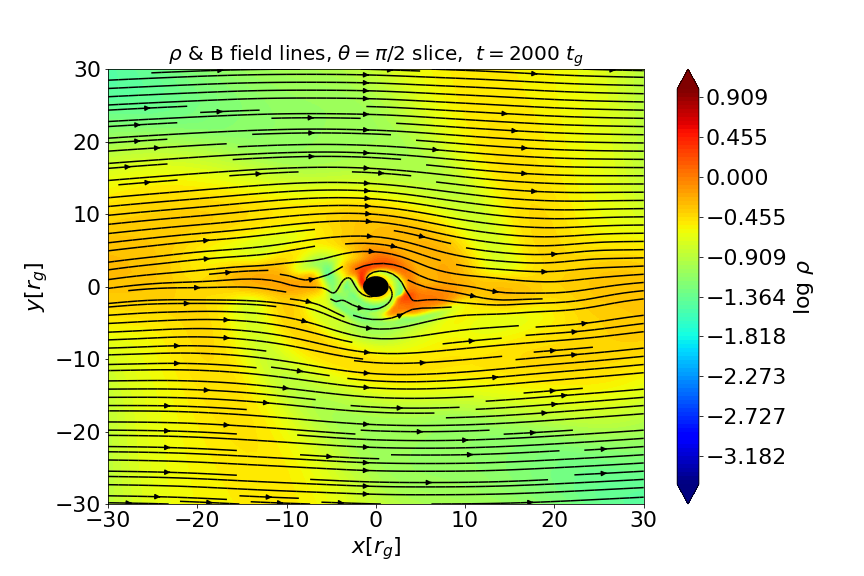}

    \includegraphics[width=0.33\textwidth]{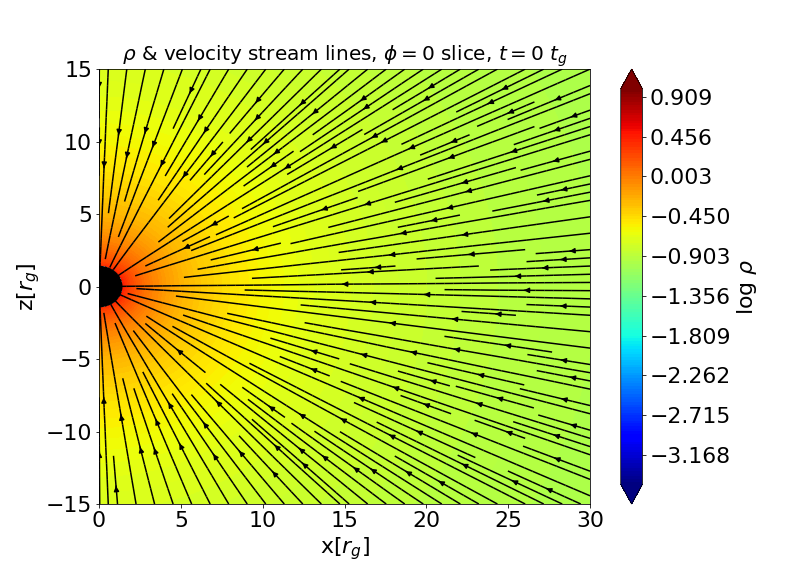}
    \includegraphics[width=0.33\textwidth]{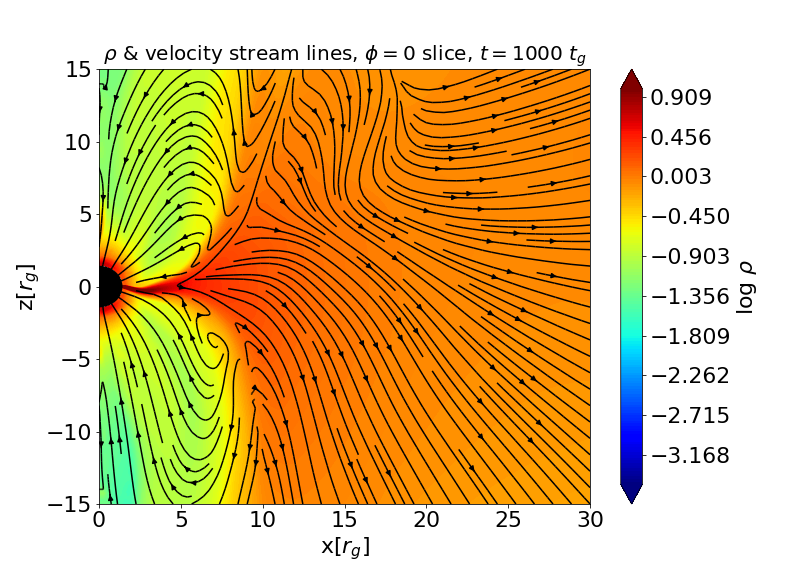}
    \includegraphics[width=0.33\textwidth]{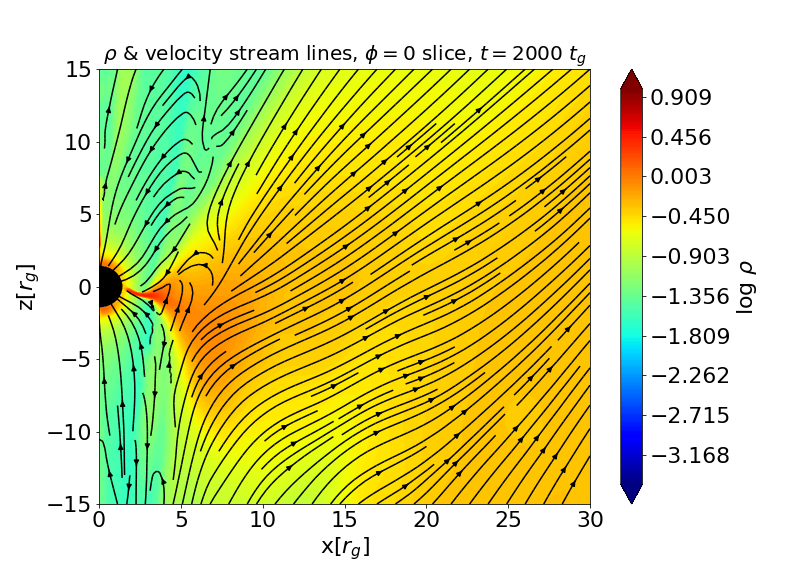}

    \includegraphics[width=0.33\textwidth]{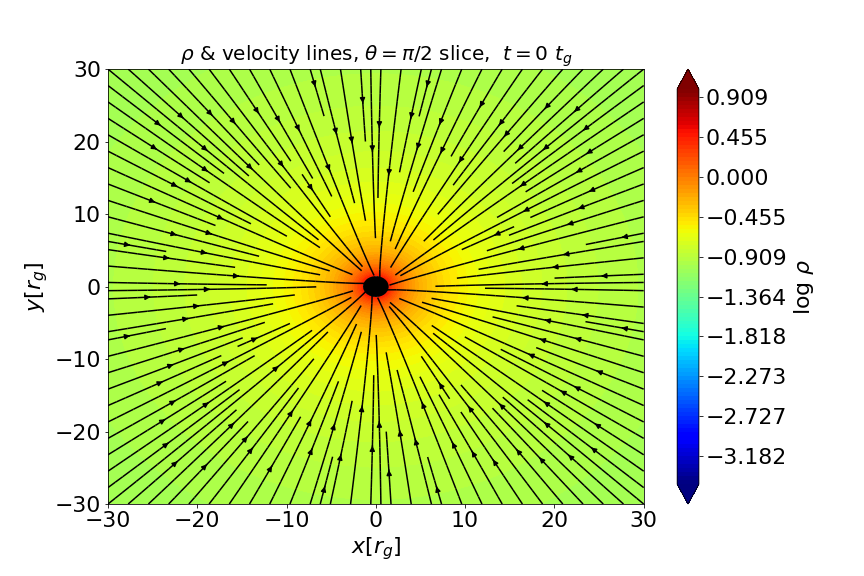}
    \includegraphics[width=0.33\textwidth]{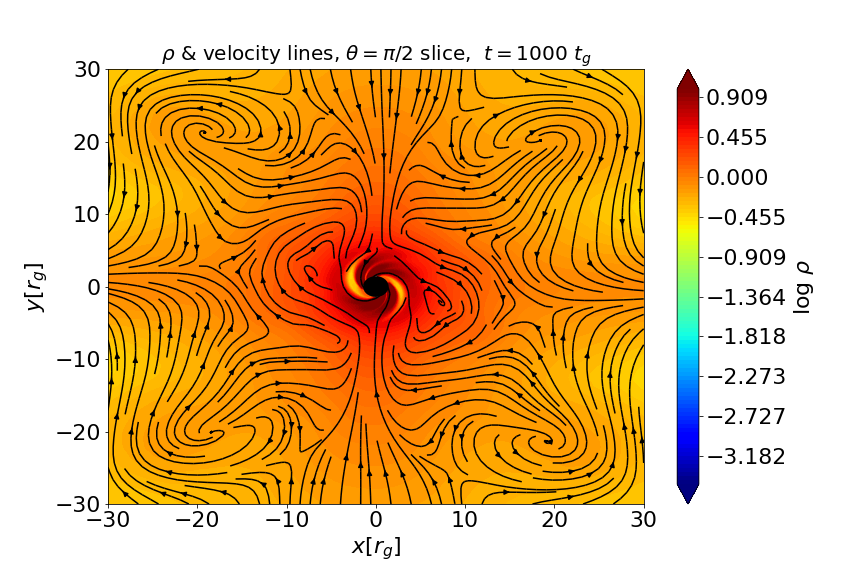}
    \includegraphics[width=0.33\textwidth]{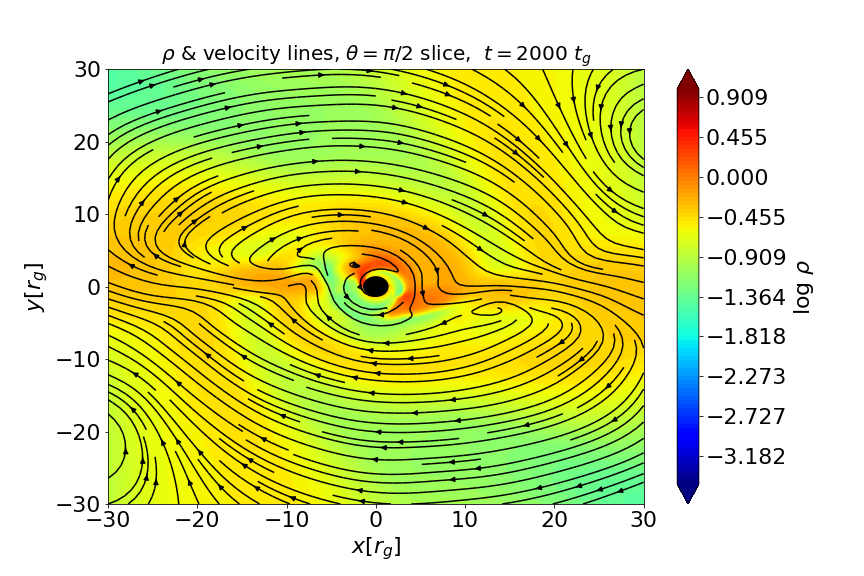}
    \caption{Selection of exemplary moments in the evolution of the 3D model with the magnetic field inclined at $45^\circ$ to the BH rotation axis exhibits density, magnetic field (top two panels), and velocity (bottom two panels) evolution of the mis-aligned configuration. 
    The left-most column shows the initial configuration when we turn on the magnetic field, while the middle and the right columns correspond to later time frames, as indicated by the labels attached to each panel. The first row from the top depicts poloidal slices at $\phi = 0^\circ$, the second row shows equatorial slices ($\theta = \pi/2$). The third and fourth rows show analogous slices with the velocity field. 
    }
    \label{fig:[inc45_a90_3D]init_evolvd_rho_B_vel}
\end{figure*}

\begin{figure}[tbh!]
    \centering
    \includegraphics[width=0.47\textwidth]{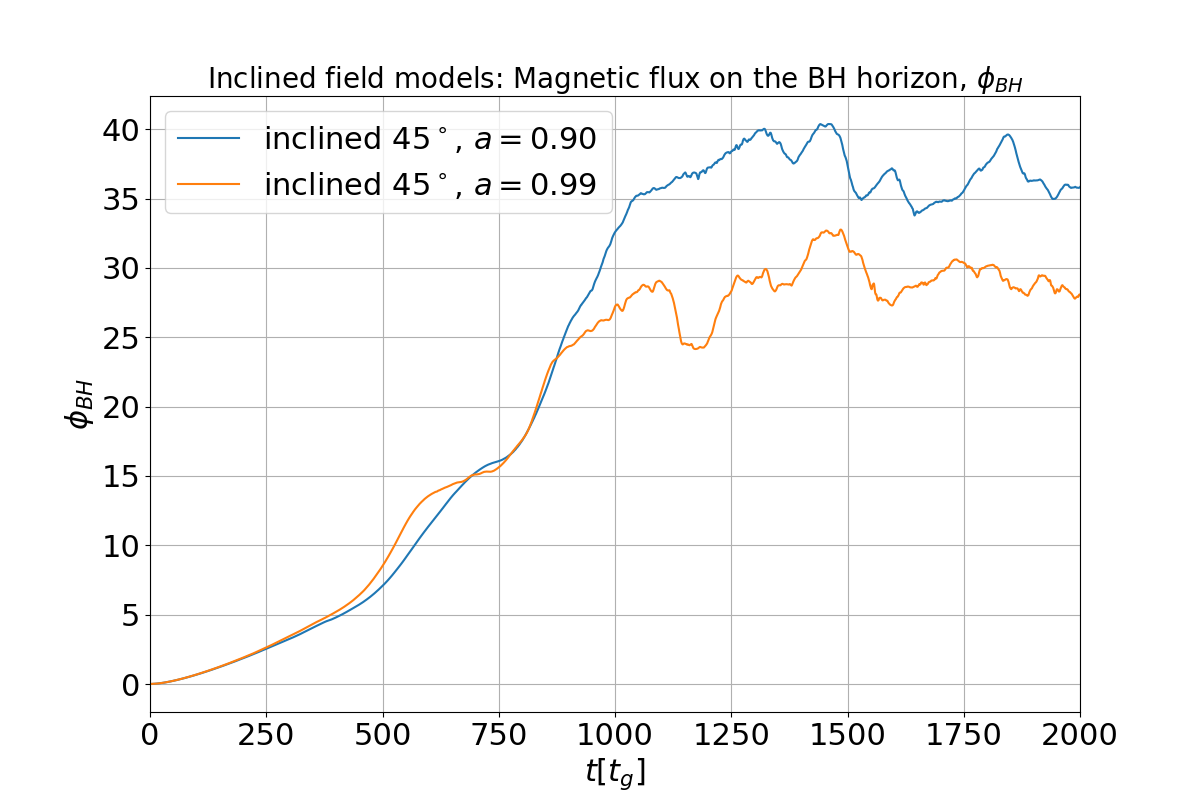}
    \caption[$\phi_\mathrm{BH}$ with time for the 3D inclined models]{The evolution of magnetic flux 
    on the black hole horizon with time for the 3D inclined field models with different spin values. 
    }
    \label{fig:inclined_phiBH_t}
\end{figure}


\begin{figure}[tbh!]
    \centering
    \includegraphics[width=0.47\textwidth]{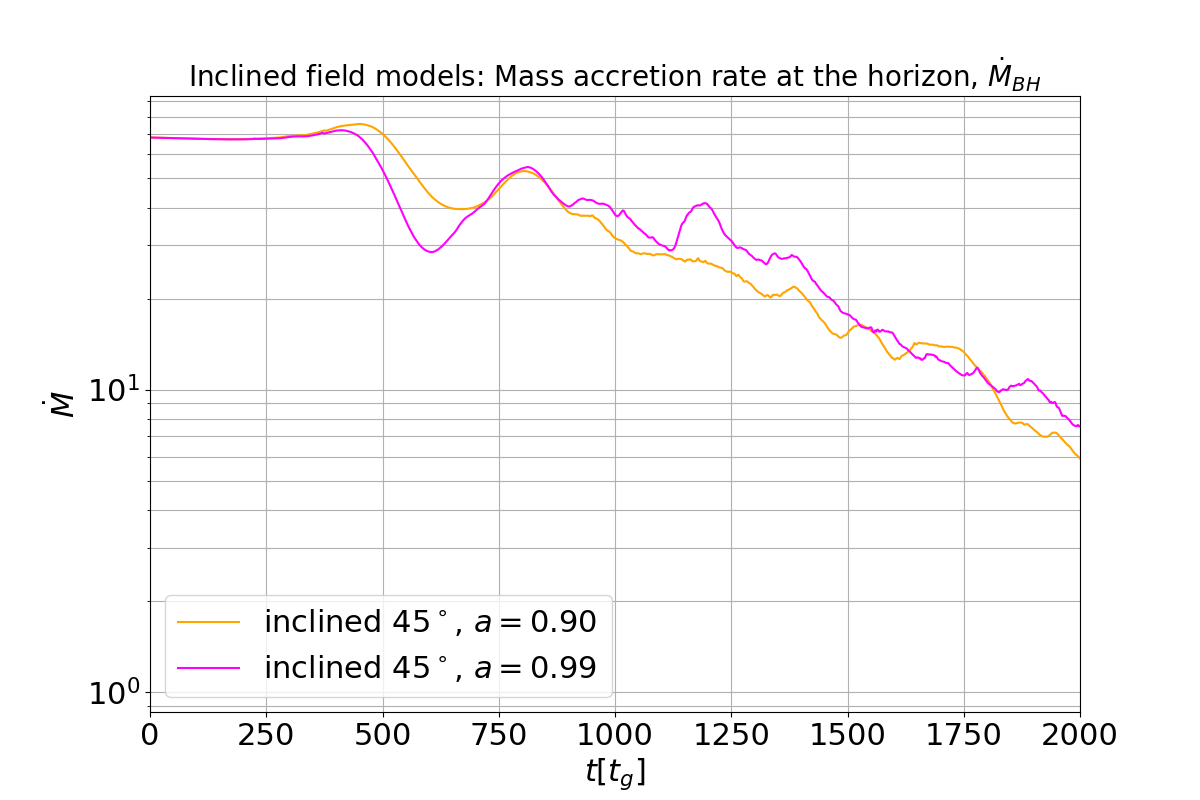}
    \caption[Mass accretion rate with time for the 3D inclined models]{The evolution of inward mass accretion rate at the black hole horizon with time, after the magnetic field is turned on, for the 3D inclined field models. }
    \label{fig:inclined_mdot_t_log}
\end{figure}

\begin{figure*}[tbh!]
    \centering
    
    \includegraphics[width=0.33\textwidth]{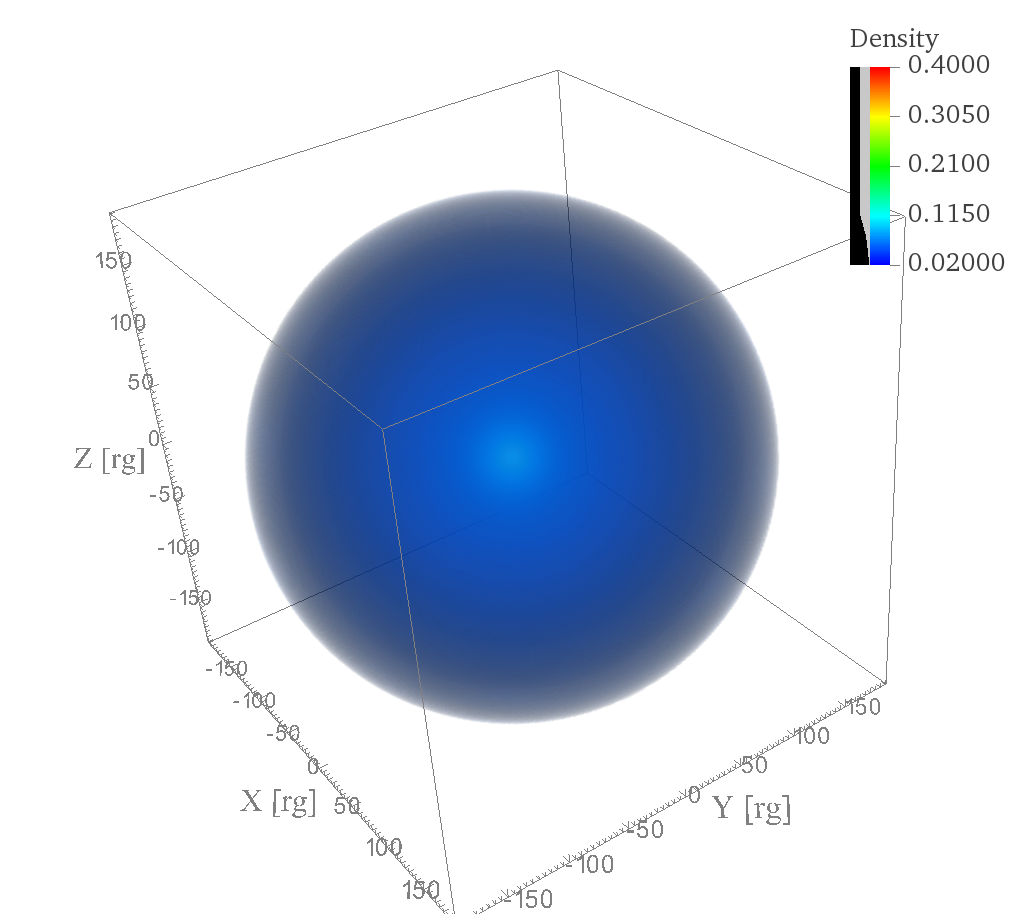}
    \includegraphics[width=0.33\textwidth]{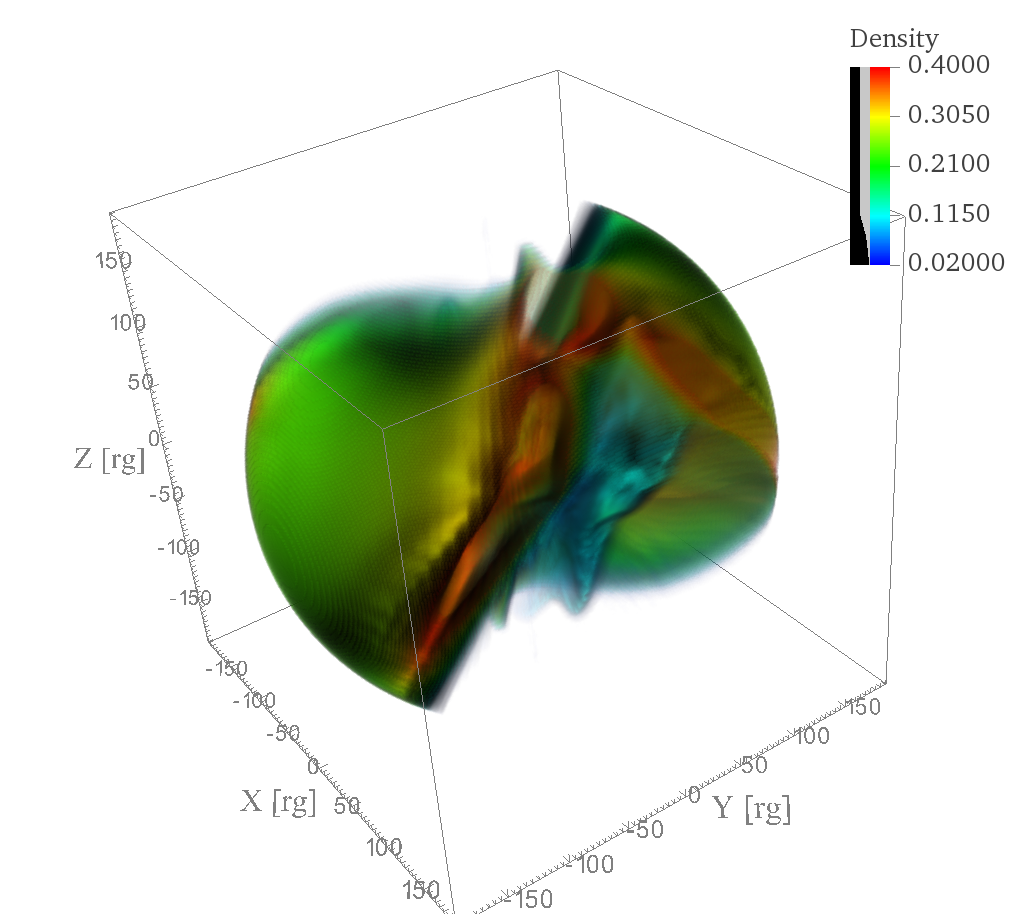}
    \includegraphics[width=0.33\textwidth]{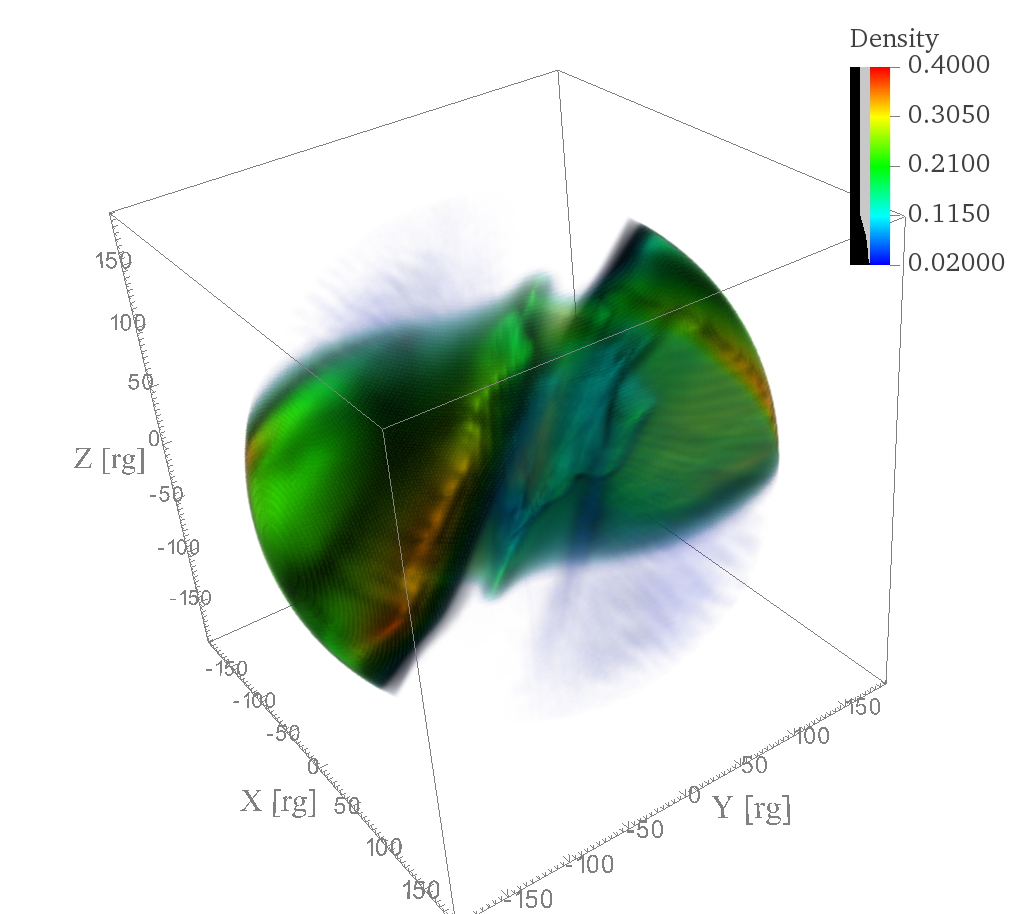}

    \caption{The volume rendered 3D plots show the initial configuration (at zero time in the left panel) and two subsequent, more evolved states (at $t=1500\,t_g$ in the middle panel, and $t=2000\,t_g$ in the right). Different color levels indicate the density for the representative run {\sf incl45.bmin0003.a90}, which covers radius up to 150\,$r_g$. Starting from the initial Bondi-type inflow at large distance, an outflow develops and it proceeds twisted and inclined with respect to the the black-hole spin (Z) axis. The color scale ranges from a negligible (close to floor) value (shown in dark blue) up to a maximum (red, in code units).}
    \label{fig:[inc45_a90_3D]rho_pcolor_visit_0}
\end{figure*}

Table \ref{tab:incl_models} gives a summary of models investigated with magnetic field inclined with the rotation axis of the black hole. The model name reflects the magnetic field normalization, the black hole spin, and the angle of inclination of the field. 
As listed in the table, we investigate a 
set of models with angle of inclination $45^\circ$ 
to the black hole rotation axis. Figure \ref{fig:[inc45_a90_3D]init_evolvd_rho_B_vel} shows the initial and evolved states of density, magnetic field and velocity field at chosen time instances for the model with magnetic field inclined at $45^\circ$. 
While the initial magnetic field strength is smaller near the horizon for these models, due to the chosen normalization, the Meissner-like expulsion of field lines are still visible in the top-left panel of Figure \ref{fig:[inc45_a90_3D]init_evolvd_rho_B_vel}. As the accretion proceeds and the plasma drags in the frozen in magnetic field lines to the black hole horizon, we can see reconnection events in the equatorial region of the flow in these models too, as similar to the aligned ones. As the simulation proceeds, we notice that the initially inclined magnetic field tends to align along the black hole rotation axis as visible in the top middle and right columns of Figure \ref{fig:[inc45_a90_3D]init_evolvd_rho_B_vel}. This in turn affects developing a more ordered and spread outflow as visible in the plots. The velocity streamlines depicted in the two bottom panels of the figure show these structured outflows originating in the equatorial region near the black horizon and eventually spreads out as they move outwards. Thus the outflows developed in these models are not confined to the equatorial region as in the case of aligned field models.

Figure \ref{fig:inclined_phiBH_t} shows the time-evolution of magnetic flux on the black hole horizon for the inclined field models. As opposed to the models with the aligned field, here we start with a field configuration that is weaker near the horizon not to disrupt the gas flow. But as accretion proceeds, the plasma brings in more magnetic flux to the black horizon and the magnetic flux on the horizon reaches very similar values ($\phi_{BH} \sim 30-40$) as compared to the aligned field models. Thus these models also reach and sustain a magnetically arrested accretion state as accretion proceeds and the inward accretion rate settles to a value that is similar to the aligned field models, as can be seen in Figure \ref{fig:inclined_mdot_t_log} and Table \ref{tab:incl_models}. Thus the models with the misaligned and aligned fields, with respect to the black hole spin, seem to reach similar inward accretion rate and similar magnetic flux on the horizon as the accretion proceeds. 
While this is true, the velocity patterns developed the in the equatorial region and the plasma distribution seem to be different when the field is misaligned with respect to the spin. This is evident when comparing the velocity stream lines plots on the equatorial region in Figures \ref{fig:[b01_a90_3D]init_evolvd_density_B_pol} and \ref{fig:[inc45_a90_3D]init_evolvd_rho_B_vel}. This difference is shown more clearly in Figure \ref{fig:vel_comparsion_al_vs_inc}. The model with the aligned field develops much more turbulent velocity patters in the equatorial region of the flow while in the inclined field model, the flow seem to be more streamlined.


\section{Discussion}
\label{diss}

\begin{figure*}[tbh!]
    \centering
    
    \includegraphics[width=0.22\textwidth]{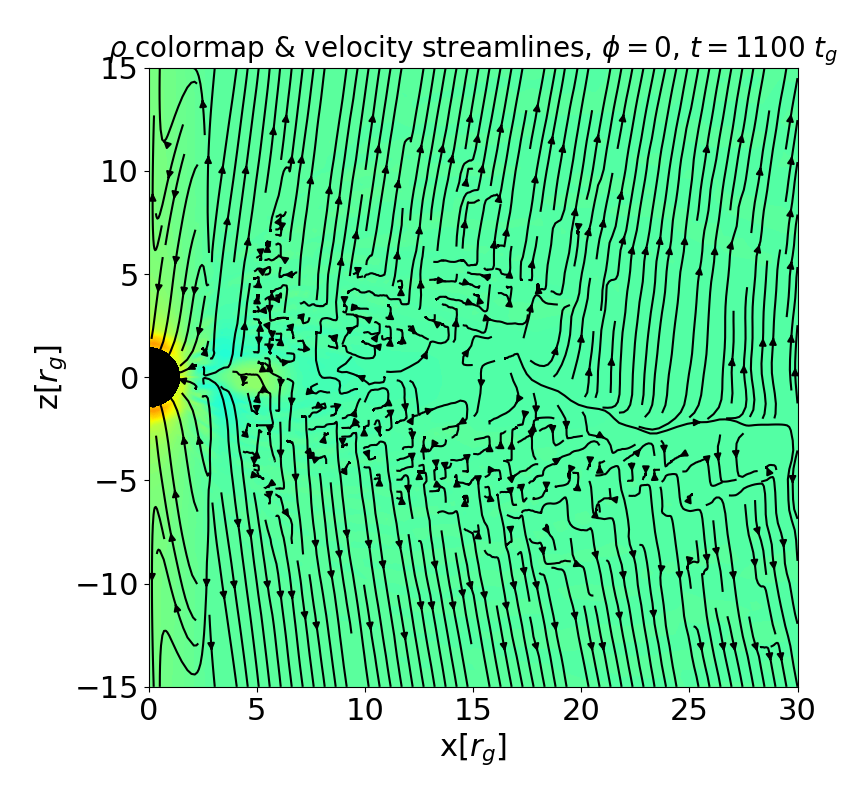}
    \includegraphics[width=0.22\textwidth]{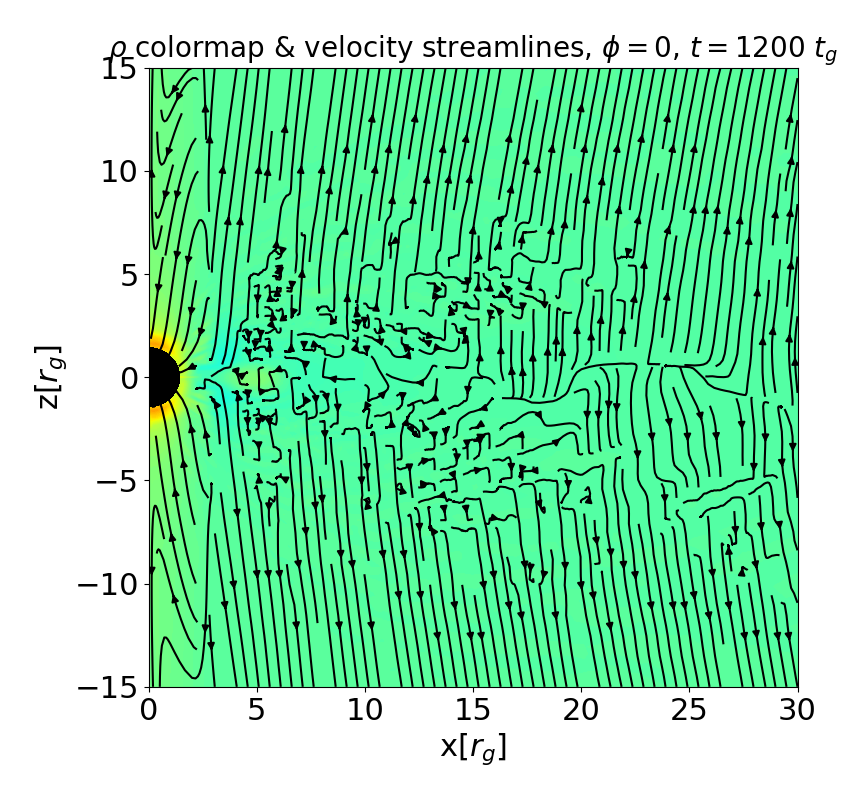}
    \includegraphics[width=0.22\textwidth]{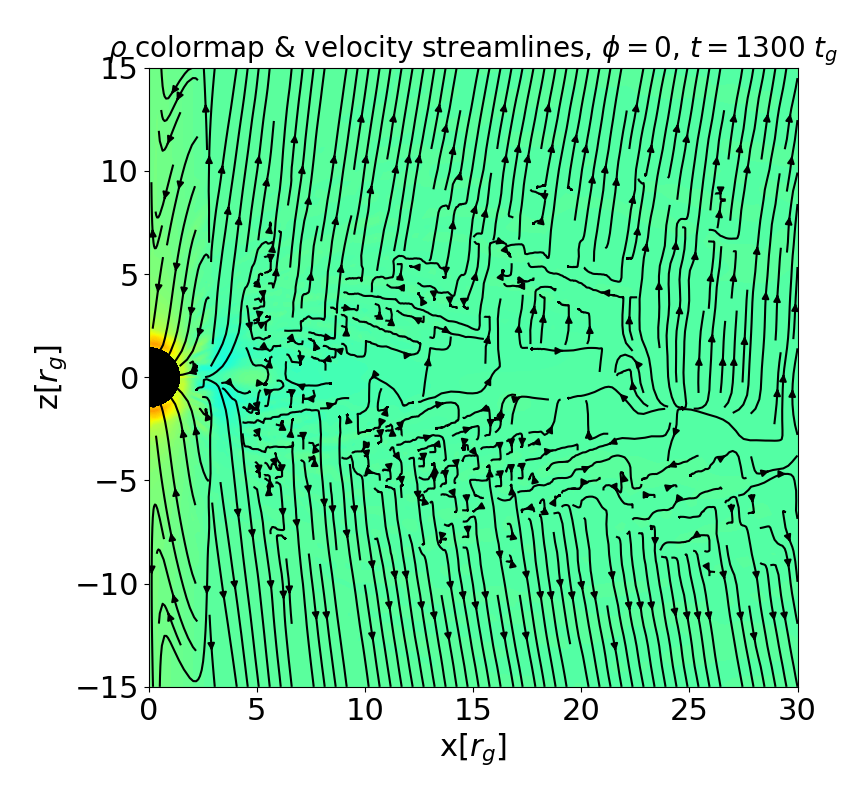}
    \includegraphics[width=0.29\textwidth]{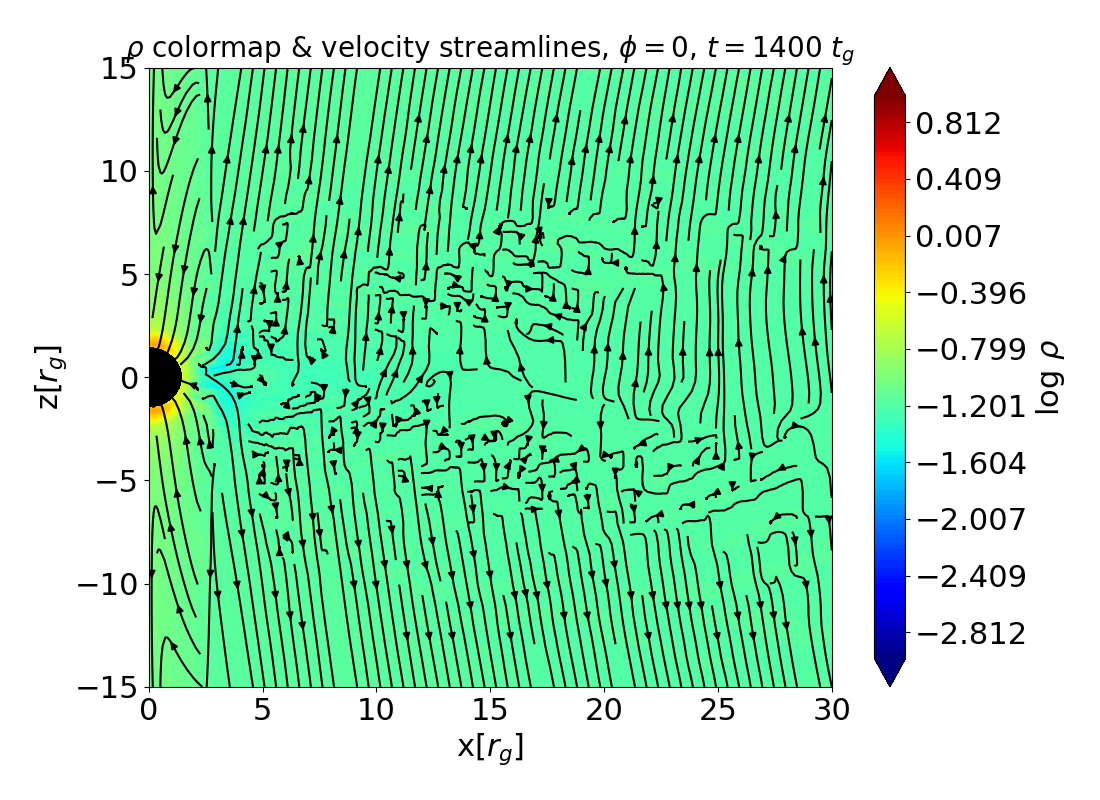}

    \includegraphics[width=0.22\textwidth]{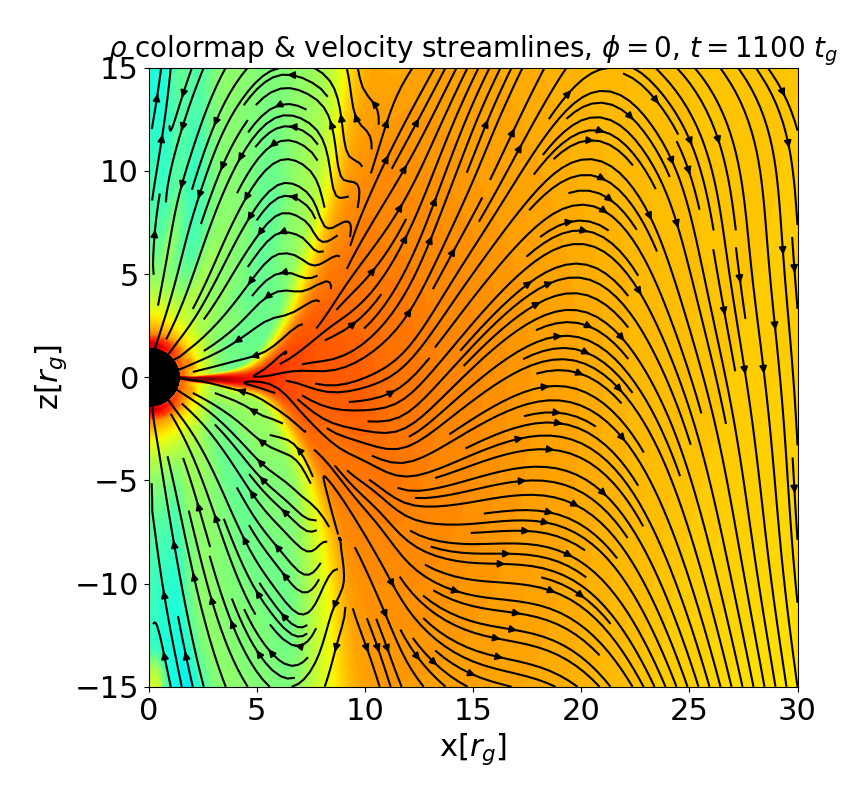}
    \includegraphics[width=0.22\textwidth]{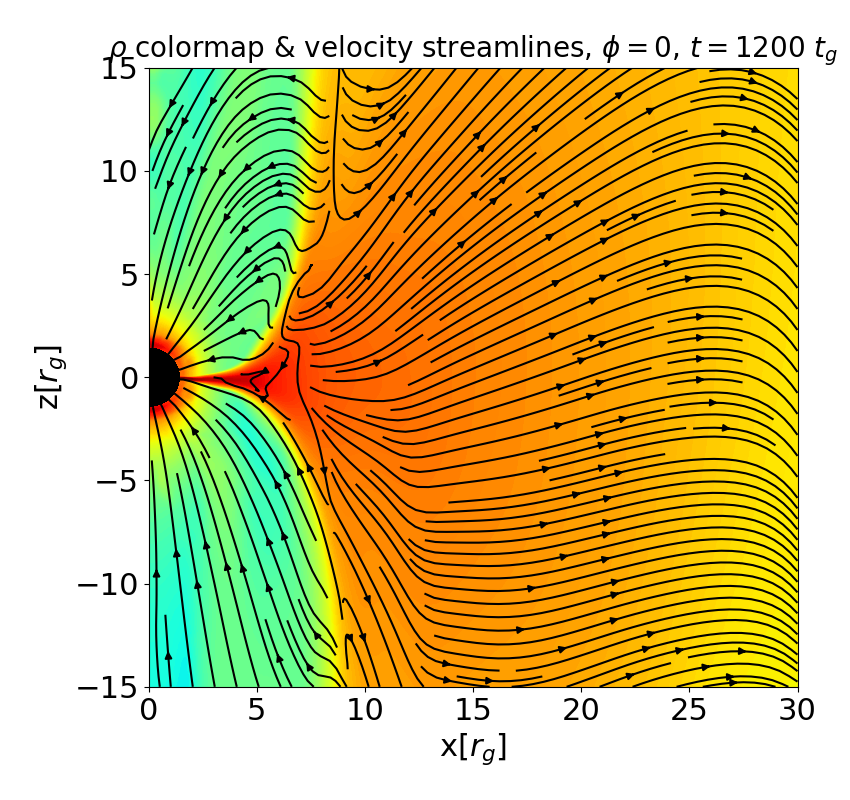}
    \includegraphics[width=0.22\textwidth]{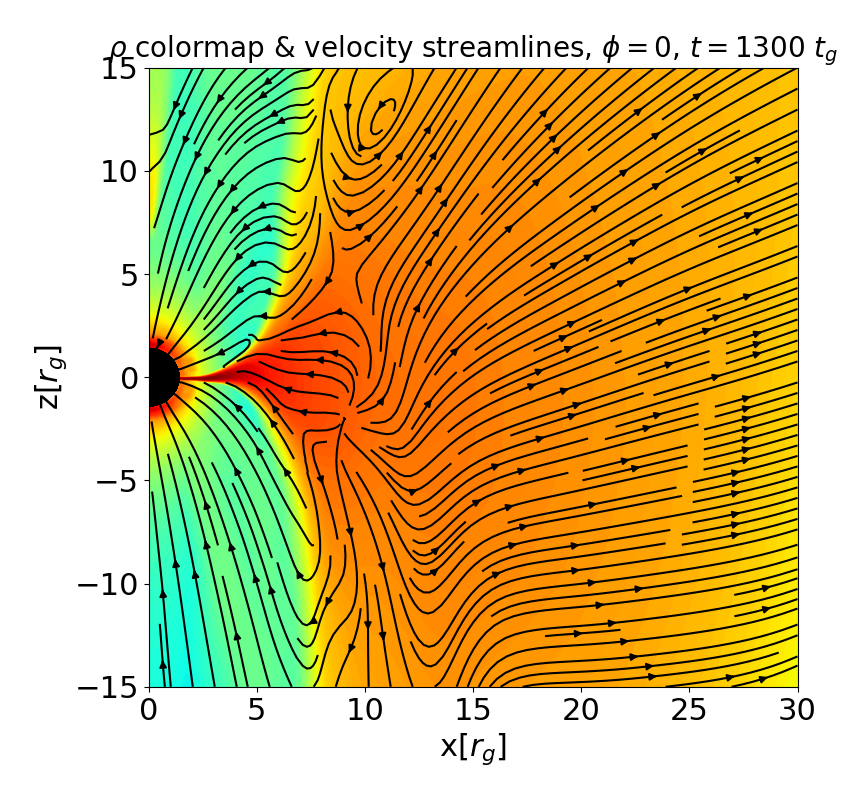}
    \includegraphics[width=0.29\textwidth]{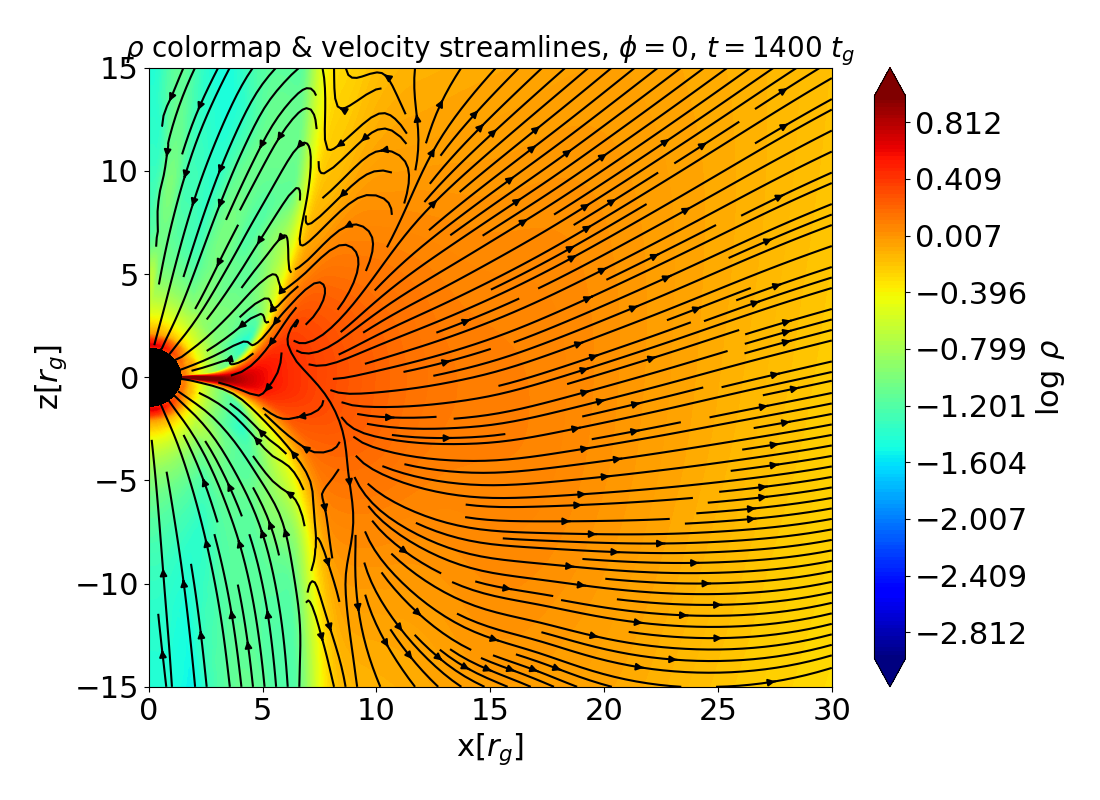}

    \caption{Comparison of velocity patterns at several time instances for the 3D aligned (b01.a90.3D) (top row) vs. 3D inclined models (incl45.bmin0003.a90) (bottom row) in the inner regions. The plots exhibit more turbulent structures developed along the equatorial region of the flow in the aligned (top) case, whereas an outflow and a low-density funnel along the spin (vertical) direction emerge in the latter (bottom) case. 
    }
    \label{fig:vel_comparsion_al_vs_inc}
\end{figure*}

There are several works investigating the inner magnetospheric evolution of black holes and its effect on the observed parameters. But, they vary in respect of their initial configuration in terms of the matter distribution, the accretion state of the system and the geometry and strength of the initial magnetic field. The configuration investigated by \citet{Ressler_2021} has similarities with our initial condition as they begin from a non-rotating inflow and an asymptotically uniform magnetic field. Their studies are focused on electromagnetic jets that are launched along the rotation axis of the black hole. We introduce the magnetic field in our models on an evolved accretion state that already reached a Bondi-like solution, and we investigate the matter outflows in the equatorial region, mediated by the formation of current sheets, and the velocity patterns there in more detail. Such a study is significant since the inner magnetopheric evolution, the matter distribution there and the associated plasmoid mediated flows there are shown to have effects on the emissions observed from such a region \citep{2022ApJ...924L..32R}. The equatorial outflow rate and the influence of magnetic field on such flows has not been investigated so for. Several kinetic simulations of the inner black hole magnetosphere, as described in the introduction, support the formation of equatorial current sheets which in turn supports the outflows and observed flares from supermassive black hole observations \citep{Crinquand2022PIC, 2022A&A...663A.169E, ElMellahetal_2023A&A}. The results from all our models models show persistent magnetic reconnection activity in the inner magnetosphere $\sim~5 - 10~r_g$, which in turn supports consistent outflows in the equatorial region. This is evident from mass outflow rate computed at $10~r_g$ for the models with the aligned magnetic field as given in Figures \ref{fig:[b01_2D]_mout_t} and \ref{fig:[b01_3D]_mout_t}. 
While being able to capture more physics at smaller scales, PIC simulations have the limitation that the initial magnetic field quickly dying out after a few tens of $t_g$ due to the current sheet getting extended to the outer boundary of the simulation box which is often provided with open boundary conditions \citep{PIC_GRBlightcurve_2021A&A}. Our models being GRMHD simulations show persistent reconnection events in the equatorial region and a continued outflow throughout the simulation timescale. Future modeling studies would benefit from a combined approach that involves a coupling between kinetic and fluid approaches, as seen e.g. in the experiments initiated by \citet{2017CoPhC.221...81Makwana}.

The 3D GRMHD models discussed in \citet{Kwan2023ApJ...946L..42K} required a minimum value of initial specific angular momentum for the models to reach and sustain a magnetically arrested state. On the other hand in our models, the flows do not posses any significant angular momentum neither at the initial nor at the evolved stages and they seem to sustain a high amount ($>20$ normalized to the inward mass flux as seen in Figures \ref{fig:[b01_2D]_phiBH_t}, and \ref{fig:[b01_3D]_phiBH_t}) of magnetic flux on the black hole horizon characteristic to a MAD state in the case of disk accretion. Moderate effects of frame dragging are visible in the inner regions of the flow where the flow and the magnetic field lines are forced to co-rotate with the black hole. This result is evident in the models with the inclined field as well, where we start with a relatively lower strength of initial magnetic field.

We note that the initial distribution of $\beta$ is not uniform in our inclined models due to the Bondi-like distribution of the gas and the resulting gas pressure. It is worth noting that in the models with the inclined field, the regions near to the black hole horizon acquires a considerable amount of magnetization with time as can be seen from Figure \ref{fig:inclined_phiBH_t} that depicts the time dependence of magnetic flux on the horizon. As the models evolved further in time, they also developed outflows, even though not specifically focused along the equatorial latitude. This can be attributed to the somewhat differently ordered distribution of the magnetic field lines. 
Thus the density patterns developed at later stages in the flow and, consequently, the outflows of matter seem to posses prevailing inclination similar to the initially imposed field geometry while also being affected by the black hole's spin (see Figure \ref{fig:[inc45_a90_3D]rho_pcolor_visit_0}). 
Also, while the general trend on the horizon (in Figure \ref{fig:inclined_phiBH_t}) shows an initial monotonic increase of the magnetic flux due accretion of the frozen-in field lines, it appears that the fluctuations caused by the flow instabilities have a magnitude comparable to the effect of spin, so the two influences cannot be disentangled.

In the models beginning with the inclined magnetic field, we also notice the formation of a low-density vertical funnel region along the rotation axis of the black hole, as opposed to no such formation in the aligned magnetic field models. This effect and the associated velocity distribution of matter is depicted in Figure \ref{fig:vel_comparsion_al_vs_inc}. 
Our models also suggest variability in density manifested in terms of variable outflows around a range of $\sim 10~r_g$. This can be relevant in the context of rapid variability from our own galactic center Sgr A*, observed within few to ten of gravitational radii above the ISCO \citep{2018A&A...618L..10GRAVITY, 2021ApJ...917...73Witzel}. While we could not verify the circular motion of plasmoids (as suggested by the GRAVITY observations), the relevance of strong poloidal magnetic fields in the innermost regions of the flow is evident from our models. The difference in the tilt of magnetic field also plays a significant effect in the turbulence developed in the flow. This can be noticed in Fig. \ref{fig:vel_comparsion_al_vs_inc} where the model with the aligned field depicted in the top row develop much more turbulent velocity patters in the equatorial region of the flow while the model with the tilted field develops more streamlined outflows.

\section{Summary and Conclusions}\label{conclusions}

We investigated the inner magnetospheric structure, the magnetic field evolution and the flow lines of plasma near the event horizon, and the associated outflows that develop in the vicinity of an accreting black hole. The details of some interesting findings can be investigated in more detail in the future; in particular, we have noticed that

   \begin{enumerate}
      \item The initially uniform magnetic field is promptly dragged by plasma inflowing into the black hole even in the case of high magnetization, $\beta\ll1$ and almost extreme spin, $a\rightarrow1$.
      \item In the aligned 2D and 3D configurations, an equatorial sheet develops and the magnetic structure suggests an enhanced turbulence and subsequent outflow that reverts a fraction of the accreted material to the outflow.
      \item Comparing 2D simulation results with the corresponding 3D models, we notice slightly increased outflow rates which can be attributed to enhanced reconnection events caught by the 3D simulations.
      \item The outflow rates have no clear dependence on the magnetic field strength while the black hole spin seems to have a more considerable effect.
      \item In the inclined field configurations, a low-density funnel develops predominantly along the spin axis rather than the initial magnetic field direction.
      \item At the same time, a dense outflow still persists at evolved stages roughly along the initial inclination of the magnetic field (in the inclined field models). 
      \item The black hole spin has a considerable effect on the flow geometry in all our models, especially in regions near the black hole horizon. This is evident in all models regardless of the magnetic field inclination. The inclination of magnetic field has effects on the flow geometry only at large distances ($\sim$ hundreds of $r_g$). 
   \end{enumerate}

\pagebreak
\begin{acknowledgements}
      We thank Dr.\ Petra Sukov\'a and Dr.\ Gerardo Urrutia for helpful discussions regarding the presentation of our work.
      We also greatly acknowledge the anonymous reviewer for very constructive and valuable comments that helped to improve our manuscript. This research was supported in part by the grant DEC-2019/35/B/ST9/04000 from the Polish National Science Center and EXPRO grant 21-06825X from the Czech Science Foundation. The Czech-Polish Mobility program of the two Academies of Sciences, titled ``Appearance and dynamics of accretion onto black holes'', is greatly appreciated. 
      We gratefully acknowledge the Polish high-performance computing infrastructure PLGrid (HPC Centers: ACK Cyfronet AGH) for providing computing facilities and support within the computational grant No. PLG/2023/016178.
      This research was carried out also with the support of the Interdisciplinary Center for Mathematical and Computational Modeling at the University of Warsaw (ICM UW) under the grant Nos. g90-1368 and g92-1500. 
\end{acknowledgements}

\bibliographystyle{aa}
\bibliography{references}

\end{document}